\documentclass[a4paper,preprint]{imsart}

\RequirePackage{amsthm,amsmath,amsfonts,amssymb}
\RequirePackage[authoryear]{natbib}
\RequirePackage[colorlinks,citecolor=blue,urlcolor=blue]{hyperref}
\RequirePackage{graphicx}

\usepackage{booktabs}
\usepackage{tabularx}
\usepackage{multirow}
\usepackage[title]{appendix}
\usepackage[ruled,lined]{algorithm2e}
\usepackage{accents}
\usepackage{mathtools}
\usepackage[dvipsnames]{xcolor}
\usepackage{luke-macros}
\usepackage[british]{babel}

\usepackage{upquote}
\usepackage{fancyvrb}
\DefineVerbatimEnvironment{verbatim}{Verbatim}{xleftmargin=15pt}

\graphicspath{{./figs/}}

\usepackage{marginnote}

\begin{document}

\begin{frontmatter}

\title{TraitLab: a Matlab package for fitting and simulating binary tree-like data}

\runtitle{TraitLab}
\runauthor{Kelly et al.}

\begin{aug}
\author[LK]{\fnms{Luke J.} \snm{Kelly}\ead[label=eLK]{lkelly@ucc.ie}\thanks{Corresponding author.}},
\author[GK]{\fnms{Geoff K.} \snm{Nicholls}\ead[label=eGK]{nicholls@stats.ox.ac.uk}},
\author[RR]{\fnms{Robin J.} \snm{Ryder}\ead[label=eRR]{ryder@ceremade.dauphine.fr}} \\
\and
\author[DW]{\fnms{David} \snm{Welch}\ead[label=eDW]{david.welch@auckland.ac.nz}}

\address[LK]{School of Mathematical Sciences, University College Cork, \printead{eLK}}
\address[GK]{Department of Statistics, University of Oxford, \printead{eGK}}
\address[RR]{CEREMADE, CNRS, UMR 7534, Université Paris-Dauphine,  Université PSL, \printead{eRR}}
\address[DW]{School of Computer Science, University of Auckland, \printead{eDW}}

\end{aug}

\begin{abstract}
TraitLab is a software package for simulating, fitting and analysing tree-like binary data under a stochastic Dollo model of evolution.
The model also allows for rate heterogeneity through catastrophes, evolutionary events where many traits are simultaneously lost while new ones arise, and borrowing, whereby traits transfer laterally between species as well as through ancestral relationships.
The core of the package is a Markov chain Monte Carlo (MCMC) sampling algorithm that enables the user to sample from the Bayesian joint posterior distribution for tree topologies, clade and root ages, and the trait loss, catastrophe and borrowing rates for a given data set.
Data can be simulated according to the fitted Dollo model or according to a number of generalized models that allow for heterogeneity in the trait loss rate, biases in the data collection process and borrowing of traits between lineages.
Coupled pairs of Markov chains can be used to diagnose MCMC mixing and convergence and to debias MCMC estimators.
The raw data, MCMC run output, and model fit can be inspected using a number of useful graphical and analytical tools provided within the package or imported into other popular analysis programs.
TraitLab is freely available and runs within the Matlab computing environment with its Statistics and Machine Learning toolbox, no other additional toolboxes are required.
\end{abstract}


\end{frontmatter}



\section{Introduction}

The ability to fit phylogenetic models to genetic sequence data has become central to many areas of the biological sciences thanks, in part, to a number of complex software tools that aid this process.
In the past two decades, phylogenetic models have also become popular for other types of data, in particular to describe the changes through time of languages and cultures. These models have been applied to many language families, including Indo-European \citep{gray2003ltd, chang2015ancestry,heggarty2023language}, Sino-Tibetan \citep{sagart2019dated}, Bantu \citep{currie2013cultural, koile2022phylogeographic}, Pama-Nyungan \citep{bouckaert2018origin} and Austronesian \citep{gray2009language}.

TraitLab\footnote{Available to download from \url{https://github.com/traitlab-mcmc/TraitLab}.} was developed specifically to fit binary trait data under a stochastic Dollo model of evolution, as described by \citet{nicholls2008}.
This is a stochastic analogue of the \citep{dollo1893} parsimony principle, in which any trait shared among individuals is assumed to have descended from the same evolutionary innovation.
Examples of data to which this model can be or has been applied are morphological traits (`has wings', `has opposable thumbs'), cultural traits (`uses curvilinear designs in woodcarving', `makes pottery') or lexical traits (`uses word with Old Saxon root \emph{al} to mean ``all''', `uses word with Latin root \emph{totus} to mean ``all''').
The Dollo assumption insists that such complex traits arise only once in the evolutionary history of the set of taxa being studied so that, for example, for a set of taxa including birds and insects, `has wings' would not be a valid trait as insect and bird wings evolved independently, but could be replaced by the valid traits `has bird wings' and `has insect wings'.
The basic Dollo model assumes that the data can be fully described by a dated tree representing the evolutionary relationships between taxa and parameters for the birth and death rates of traits.

An extended Dollo model, also implemented in TraitLab, allows so-called catastrophes to occur in which multiple traits are born and die simultaneously, representing rapid evolutionary bursts.
This allows for heterogeneity in the rate of trait evolution along different lineages.
The extended model introduces two further parameters for the rate at which catastrophes occur and the probability of trait death at a catastrophe.
TraitLab also allows for borrowing in the Dollo model, whereby species acquire traits from contemporary species outside of ancestral relationships, which introduces a parameter for the rate at which each trait instance attempts to transfer laterally.

TraitLab fits the model within a Bayesian framework using Markov chain Monte Carlo (MCMC) techniques to draw samples from the posterior distribution of the parameters for given data.
Any or all of the parameters --- the dated binary tree, the trait death and transfer rates, the catastrophe occurrence rate and the trait death rate --- can be estimated or fixed.
The program is controlled via a simple graphical user interface in the Matlab computing environment or via a configuration file specified from the command line.

The data we consider consist of $N$ traits that have been recorded as present, absent or missing (presence or absence undetermined) in $L$ taxa.
The data $D$ are therefore an $L \times N$ binary matrix where the $(i,j)$th entry $D_{ij} = 1$ if the $j$th trait is present in the $i$th taxon, $D_{ij} = 0$ if it is absent, and $D_{ij} = ?$ if its status is undetermined.
In addition, clade constraints that place constraints on the topology and ages of certain parts of the tree can also be handled by TraitLab and are discussed in \secref{sec:cladeblock}.

TraitLab can be used to infer phylogenies from any binary data for which the stochastic Dollo model is relevant, but the majority of uses of TraitLab so far have been for inferring or simulating language phylogenies of cognate lexical data \citep{atkinson2005wdw, greenhill2009does, Koch2016, atkinson2006species, greenhill2020bayesian, sagart2019dated}, and this is the use case we consider in \appref{app:semitic}.
Applications of the stochastic Dollo model in other fields include miRNA evolution \citep{thomson2014critical} and cancer biology \citep{mcpherson2016divergent}.

Several other packages exist for phylogenetic analysis of binary data, with significantly different assumptions from TraitLab.
For instance, MrBayes \citep{ronquist12mrbayes} implements the finite-sites trait evolution model of \citet{lewis2001lae}, which assumes that the number of traits is fixed in advance, and that a trait can be born independently at several points on the tree, whereas the stochastic Dollo model we fit allows for the number of traits to be random, but each trait can only be born once (homoplasy is not allowed).
MrBayes has nonetheless also been applied to lexical traits \citep{gray2003ltd}.
Similarly, BayesPhylogenies \citep{pagel2004bayesian} also implements MCMC for a phylogenetic model in which a trait can be born at several different points.
BEAST \citep{suchard2018beast1} proposes MCMC implementations of several models, including the related multi-state stochastic Dollo process of \citet{alekseyenko2008wad}, but without our modelling extensions, and without the possibilities included in TraitLab includes for simulations of synthetic data from generalized models.
\citet{barbancon2013} compare several packages for the analysis of lexical trait data. Beast 2 \citep{bouckaert2019beast2} also has an implementation of the stochastic Dollo model, and of the related pseudo-Dollo model \citep{bouckaert2017pseudo}. To our knowledge, TraitLab is one of only two implementations of a model with horizontal transfer of lexical items, the other being \texttt{contacTrees} \citep{neureiter2022detecting}.

Many ad hoc approaches exist which attempt to diagnose MCMC convergence in phylogenetics problems but may fail silently.
A coupling of the MCMC algorithm, whereby a pair of coupled Markov chains explore the target posterior distribution and meet exactly after a random number iterations, can be used to jointly diagnose convergence across all components of the model and debias estimators constructed from MCMC samples.
At the time of writing, TraitLab is the only phylogenetics software to implement this coupling approach.

Further details on the basic model, its implementation and its application to linguistic data can be found in \citet{nicholls2008}, \citet{atkinson2005wdw} and \citet{nicholls2011phylogenetic}. The extension for rate heterogeneity through catastrophes is described in \citet{ryder2011missing} with additional detail in \citet{ryder2010dphil}. Incorporating lateral trait transfer in the stochastic Dollo model is described \citet{kelly2017lateral} with further details contained in \citet{kelly2016dphil}.
The algorithm to couple MCMC for phylogenetic inference is described by \citet{kelly2023lagged}.

The remainder of the paper is organized as follows.
Sections~\ref{sec:model}--\ref{sec:model-extensions} describe the trait evolution and data collection models.
\secref{sec:prior} outlines the prior distributions and \secref{sec:post} describes the corresponding posterior.
\secref{sec:mcmc} discusses the construction of the MCMC algorithm for sampling from the posterior.
Directions on how to install and start TraitLab are given in \secref{sec:install}.
\secref{sec:data} specifies the data file format.
Sections~\ref{sec:runmcmc}-\ref{sec:run-coupling} give step-by-step instructions for running an MCMC analysis while \secref{sec:output} describes the tools provided for analysing and visualising the output of the MCMC run.
Finally, in \secref{sec:synth}, we describe tools to simulate data under the Dollo model and various extensions, and how this synthetic data can be used to check for model fit and model misspecification.
\appref{app:semitic} describes a step-by-step analysis of lexical data from Semitic languages.

\section{Methods}

\subsection{Notation}
\begin{figure}
    \centering
    \includegraphics[width=\textwidth]{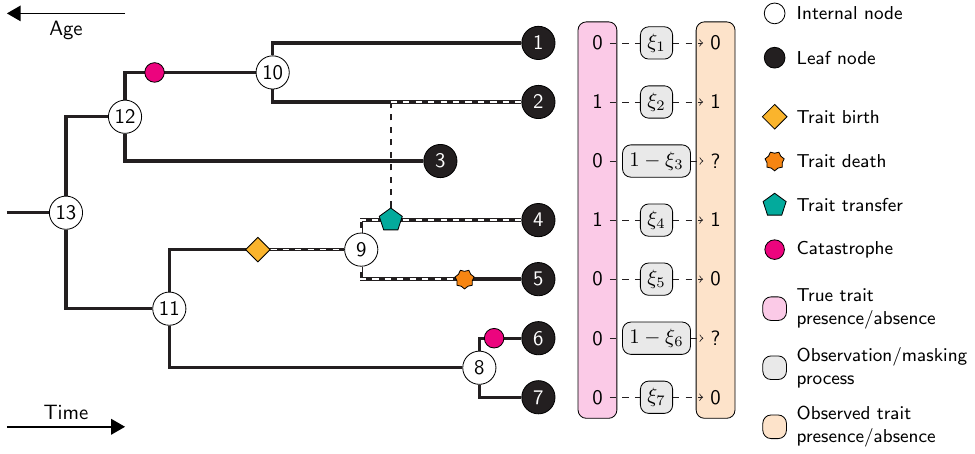}
    \caption{
        A phylogeny with $ L = 7 $ leaves and a trait history sampled from the stochastic Dollo model.
        The leaves correspond to the observed taxa, the indexing of the internal nodes is completely arbitrary.
        The root is node 13.
        The trait born on edge 9 was copied into the offspring lineages 4 and 5, was borrowed into edge 2 from 4 and died out on 5.
        The trait was present at leaves 2 and 4 and absent at the rest but, due to a missing-at-random masking process, its status was not recorded in leaves 3 and 6.
        The trait in this example did not pass through either catastrophe node on the tree.
    }
    \label{fig:phylogeny-dollo}
\end{figure}

Let $g$ be a binary rooted tree with $2L-1$ nodes: $L$ leaves and $L-1$ internal nodes, of which the \emph{root} is the eldest and the most recent common ancestor of all the leaves.
Let $V = \{1, \dotsc, 2L - 1\}$ be the set of node labels; in particular, the leaves bear the labels $ 1, \dotsc, L $.
Let $ t_i $ denote the age of node $ i \in V $ and $t = (t_1, \dotsc, t_{2L - 1})$.
Node ages increase along paths from the leaves towards the root of the tree.
The directed edges (also called branches) of $g$ run forwards in time from a parent node to its child. We index edges by their offspring node so refer to the edge leading into node $ i $ from its parent as edge $ i $.
Additionally, an edge of infinite length leads into the root from its parent, but we ignore the parent node in our notation.
Let $ E $ denote the set of edges in $ g $. Thus, $\abs{E} = 2L - 1$ as it consists of the $2L-2$ internal edges of $g$ and the edge of infinite length leading into the root node from its parent.
The tree is thus defined by $g = (V,E,t)$.
\figref{fig:phylogeny-dollo} displays a rooted, binary tree.

\subsection{A stochastic Dollo model of evolution}\label{sec:model}

In the basic Dollo model of species diversification, three types of events may occur: traits are born, traits die, and traits are duplicated when the lineages containing them split.
All evolutionary events are independent, so traits do not interact within or across species and speciation events occur independently of trait events.
Traits are born in each species according to a Poisson process with constant rate $\lambda$ in each species.
Each trait instance in each lineage dies at constant per capita rate $\mu$.
When a lineage $i$ splits, it is replaced by two identical copies of itself, $ j $ and $ k $, say, which continue to evolve independently of all other lineages.
The trait process is in equilibrium at the root, since the edge leading into has infinite length, then evolves along the tree to the leaves where the presence or absence of each trait is recorded.
\figref{fig:phylogeny-dollo} displays the history of a trait drawn from the stochastic Dollo model including the extensions described in \secref{sec:model-extensions}.

\subsection{Likelihood}

To calculate the likelihood of data recorded at the leaves given the tree and model parameters, we must integrate over the unobserved evolutionary events on the phylogeny under the model.
Fix an ordering of the leaf nodes.
Each fully observed trait displays a binary pattern of presence-absence across the leaves. For example, the trait in \figref{fig:phylogeny-dollo} would display the pattern $p=(0,1,0,1,0,0,0)$ if it were fully observed.
We cannot observe a trait absent in all of the leaves so do not consider the pattern $ (0, \dotsc, 0) $ so the space of observable patterns is $ \cP = \{0, 1\}^L \setminus \{(0, \dotsc, 0)\}$.

Let $ N_p $ denote the number of traits in $ D $ displaying pattern $ p \in \cP $ at the leaves and denote $ \EE[N_p \mid g, \lambda, \mu] $ its expected frequency under the model.
We have a Poisson process of trait births with independent thinning so $ N_p \mid g, \lambda, \mu \sim \Pois{\lambda z_p} $, where $ z_p = \EE[N_p \mid g, \lambda = 1, \mu] $.
Thus, in the basic stochastic Dollo model the likelihood of the pattern counts given the parameters is
\[
    \pi\big((N_p : p \in \cP) \mid g, \lambda, \mu\big)
        = \prod_{p \in \cP} \frac{(\lambda z_p)^{N_p} e^{-\lambda z_p}}{N_p !}
        \propto \lambda^{\sum_{p \in \cP} N_p} e^{-\lambda \sum_{p \in \cP} z_p} \prod_{p \in \cP} z_p^{N_p}.
\]
We place an improper prior $ \pi(\lambda) \propto 1 / \lambda $ on the birth rate, then we integrate it out of the prior-weighted likelihood analytically to obtain
\begin{align}
    \label{eq:int-like}
    \pi\big((N_p : p \in \cP) \mid g, \mu\big)
        &= \int_0^{\infty} \pi\big((N_p : p \in \cP) \mid g, \lambda, \mu\big) \pi(\lambda) \ud \lambda \nonumber\\
        &\propto \prod_{p \in \cP} \left(\frac{z_p}{\sum_{q \in \cP} z_q}\right)^{N_p},
\end{align}
a multinomial distribution across the patterns conditional on the total number of traits observed.
See \citet{nicholls2008} for further details on the model and likelihood computation.

\subsection{Model extensions} \label{sec:model-extensions}

\subsubsection{Catastrophic rate heterogeneity}

Catastrophes are added to this model to account for rate heterogeneity along edges.
Catastrophes occur independently along each lineage at rate $\rho$.
At a catastrophe, each trait dies with probability $\kappa$ and a $ \Pois{\nu} $ number of new traits are born.
We impose the condition that $ \lambda / \mu = \nu / \kappa $ so that the expected number of traits in any lineage remains constant over time and the process is reversible.
This is equivalent to instantaneously advancing the trait process at each catastrophe on the edge by $ -\mu^{-1} \log(1 - \kappa) $ units of time: it is therefore convenient to extend the definition of a tree to include the number of catastrophes that occur on each edge.
Denote $ k_i $ the number of catastrophes that occur on branch $ i $ and $ k = (k_1, \dotsc, k_{2L - 2}) $ the vector of catastrophe counts for all finite edges.
The extended tree is defined by $g = (V, E, t, k) $ and the parameter space is also expanded to include the catastrophe strength $ \kappa $.
See \citet{ryder2011missing} for further details of the catastrophe process and how to incorporate it in the likelihood calculation.

\subsubsection{Observation model for traits}\label{sec:obs}

Some kinds of traits are deliberately removed from the data.
For example, in the absence of lateral trait transfer, it is common to discard traits displayed at just one taxon. On the other hand, when lateral trait transfer is suspected, singleton traits are evidence of a lack of borrowing.
If only traits in observed in at least $ d $ taxa are registered in the data (with $ d = 1 $ or $ d = 2 $), then we restrict $ \cP $ accordingly and make the corresponding marginalisation of the likelihood exactly.
Secondly, some data may be missing from the matrix where it has not been possible to record the presence or absence of a particular trait in a taxon.
We model each trait as missing uniformly at random within each taxon, so that each trait is missing independently from taxon $i$ with probability $1 - \xi_i$ or else recorded with probability $\xi_i$.
When missing data is allowed in the model, the parameter space is extended to include $ \xi = (\xi_1, \dotsc, \xi_L) $.
See \citet{ryder2010dphil} and \citet{ryder2011missing} for further details of these extensions for missingness and data registration operations.

\subsubsection{Borrowing}

To account for borrowing of traits between evolving species, we allow for a global lateral transfer process whereby each trait-instance in each lineage attempts to transfer a copy of itself to other branches at rate $ \beta $.
An attempted borrowing succeeds if the trait is not already present on the destination branch.
When lateral transfer occurs, the trait process is no longer time-reversible since the rate at which traits transfer depends on the number of instances of each trait and the number of edges.
In this setting, a catastrophe advances the trait process along the affected edge by $ -\mu^{-1} \log(1 - \kappa) $ units of time relative to the other edges, and during a catastrophe traits may be born, die or transfer into the affected edge but cannot transfer out.

When the model allows for lateral transfer, the parameter space is extended to include $ \beta $.
Furthermore, if catastrophes are allowed in the model, then we expand the tree $ g $ to include their locations along edges.
See \citet{kelly2017lateral} for a complete description of lateral transfer in the stochastic Dollo model accounting for catastrophes, missing data and data registration, and \citet{kelly2016dphil} for a detailed discussion of the computation.

\subsection{Priors} \label{sec:prior}

TraitLab operates in the subjective Bayes framework. This puts an onus on the user to specify priors encoding the prior knowledge which is actually available. The following priors may need to be adjusted for any given application. Those given below may be appropriate for analysis of lexical trait data, in which age is measured in units of years.

Our prior on the death rate $ \mu \sim \Gamma(0.001, 0.001) $, with mean 1 and variance 1000, and likewise for the borrowing rate $ \beta $.
These broad priors were chosen to represent a lack of prior knowledge.
The user may change these priors by editing \texttt{LogPriorParm.m}
The prior on the catastrophe rate $ \rho $ is $ \Gamma(1.5, 5000) $.
The parameters of this Gamma prior were chosen to place 90\% prior mass on an interval of 1300 and 28000 years between catastrophes.
The user may change this prior by editing \texttt{LogRhoPrior.m}.
So that the effects of catastrophes are identifiable with respect to the underlying process, we place a $ \Unif{0.25, 1} $ prior on $ \kappa $, the death probability at a catastrophe, thus ruling out weak catastrophes.
We take an independent $ \Unif{0, 1} $ prior for each missing data parameter $\xi_i$.
Two classes of priors on trees are available.
One, which we call an exponential prior, is an exponential branching process with rate $ \theta $ stopped at the instant of the \emph{L}th branching event (counting the branching at the root as the first event). This determines a distribution
\begin{equation}
    \label{eq:exponential_tree_prior}
    f_E(g | \theta) \propto \theta^{L-1} \exp(-\theta \abs{g})
\end{equation}
where $ |g| $ is the sum of all branch lengths beneath the root.
The scale parameter $ \theta $ is given a conjugate Jeffreys prior $1 / \theta$ then integrated out, yielding the marginal tree prior $ f_E(g) \propto \abs{g}^{-L + 1} $.

The other prior on the tree $g$ is chosen in such a way that the marginal prior on the root age is approximately uniform over the interval $ [s, T] $, where $ s $ is the maximum of the lower bounds on node ages imposed by the clade constraints (or zero of there are no clade constraints) and $T$ is a maximum root age imposed by the user.
\citet{nicholls2008} show that the prior distribution with density
\begin{equation}
\label{eq:unif_tree_prior}
    f_G(g \mid T)
        \propto \ind{t_r < T} \prod_{i \in F(g) \setminus \{r\}} \frac{T - s_i^-(g)}{t_r - s_i^-(g)}
\end{equation}
where $ F(g) $ is the set of \emph{free} nodes in $ g $ with ages upper bounded only by $T$ and $ s_i^-(g) $ is the least age node $ i $ can take on the tree $ g $, induces an approximately uniform marginal prior distribution over the root age, provided that the greatest upper bound imposed by the clade constraints is not too close to $ T $.
See \citet{nicholls2011phylogenetic} for further details of this tree prior.
When clade constraints are placed on the tree, as described in \secref{sec:cladeblock}, the prior is defined to be zero for any tree not satisfying those constraints.

\subsection{Posterior distribution} \label{sec:post}

We multiply the expression for the likelihood \eqref{eq:int-like} by the corresponding priors to obtain the posterior density for the model, with parameters on their valid ranges and trees satisfying all of the clade constraints and with root age less than $ T $.
TraitLab determines the likelihood and requisite priors from the model specification.
Similar to the birth rate $ \lambda $ in \eqref{eq:int-like}, we integrate out the catastrophe rate $ \rho $ to obtain a Negative Multinomial distribution for catastrophe counts along branches.
If desired, $ \rho $ can be fixed at a constant value and included in the model.
See \citet[Supplement A]{kelly2023lagged} for further details of these properties.

\subsection{MCMC algorithm} \label{sec:mcmc}

Let $ x $ denote the tree and parameters to infer for the chosen model after marginalisation of $\lambda$ and $\rho$, which ranges from $ x = ((E, V, t), \mu) $ for the basic model to $ x = ((E, V, t, k), \mu, \beta, \kappa, \xi) $ for the full model including lateral trait transfer, catastrophes and missing data.
The target of our inference is the posterior distribution $ \pi(x \mid D) $,
We sample from the posterior distribution using MCMC.
TraitLab uses the Metropolis--Rosenbluth--Teller--Hastings algorithm \citep{metropolis1953equation,hastings70monte} to construct an ergodic Markov chain $ X_0, X_1, \dotsc $ whose limiting distribution is $ \pi(x \mid D) $, where a state of the Markov chain is some value for each component of $ x $.

General updates to the state $ x $ are intractable, so our MCMC uses a mixture of local proposal kernels, where each kernel targets a different component of the state.
Brief descriptions of the different mechanisms used to propose new states in the chain are listed in \tabref{tab:moves}.
There are four proposals that alter the tree topology and which are described by \citet{drummond2002}, five proposals that alter the heights of some or all nodes in the tree, five proposals that add, delete or shift catastrophes and which are described in \citet{ryder2011missing} and \citet{kelly2023lagged}, while the remaining five moves multiply one or more of the scalar parameters $ \mu $, $ \beta $, $ \kappa $ and $ \xi = (\xi_1, \dotsc, \xi_L) $ by a randomly chosen factor.
See \citet[Supplement A]{kelly2023lagged} for a recent description of these moves.

\begin{table}
    \centering
    \caption{
        The moves currently used in the MCMC sampler to explore the state space.
        Move indices correspond to their indices within TraitLab.
        Some moves that were used in earlier versions of TraitLab have been removed.
    } \label{tab:moves}
    \begin{tabular}{@{}lcl@{}}
        \toprule
        Primary target & Move & Description \\ \midrule
        \multirow{4}{*}{Topology}
            & 2 & Interchange parents of neighbouring node pair \\
            & 3 & Interchange parents of randomly chosen node pair \\
            & 4 & Move edge to neighbouring location \\
            & 5 & Move edge to randomly chosen location \\ \midrule
        \multirow{5}{*}{Node times}
            & 1 & Vary internal node time\\
            & 11 & Vary leaf time \\
            & 6 & Rescale tree \\
            & 7 & Rescale subtree \\
            & 12 & Rescale tree above clade bounds \\ \midrule
        \multirow{5}{*}{Catastrophes}
            & 13 & Add catastrophe \\
            & 14 & Delete a catastrophe \\
            & 17 & Move catastrophe on edge \\
            & 18 & Move catastrophe to neighbouring edge \\
            & 15 & Resample all catastrophes on edge \\ \midrule
        \multirow{5}{*}{Model parameters}
            & 8 & Rescale death rate $ \mu $ \\
            & 21 & Rescale transfer rate $ \beta $ \\
            & 16 & Rescale catastrophe strength $ \kappa $ \\
            & 19 & Rescale one missing data parameter $ \xi_i $ \\
            & 20 & Rescale all missing data parameters $ \xi $ \\
        \bottomrule
    \end{tabular}
\end{table}

\subsection{Coupling}
\label{sec:coupling}

Following the framework developed by \citet{jacob20} and \citet{biswas19}, TraitLab allows the user to sample a pair of lag-$ l $ coupled Markov chains, $ (X_s)_{s \geq 0} $ and $ (Y_s)_{s \geq 0} $, which can be used to estimate 
convergence bounds and debias MCMC estimators.
In short, having first sampled $ X_0, \dotsc, X_l $ from the marginal MCMC kernel targeting $ \pi(x \mid D) $, the pair $ (X_s, Y_{s - l}) $ are drawn from a coupling of their marginal transition kernels at each iteration $ s > l $.
The chains meet exactly at a random finite time $ \tau^{(l)} $ and remain together thereafter.
Provided that the tail of $ \tau^{(l)} $ tail decays, at most, geometrically then we can use the meeting times to estimate a bound on the total variation distance between the marginal distribution of $ X_s $ and the target $ \pi(x \mid D) $.
\citet{kelly2023lagged} describe the algorithm in detail and illustrate its utility in diagnosing convergence across all components of the phylogenetic model from the meeting times of pairs of chains run in parallel.
Coupling may also be used to check for overparameterization; for example, if catastrophes are only weakly identified in the model then coupled chains will struggle to meet.
\secref{sec:run-coupling} describes how to apply the coupling scheme in TraitLab.

\subsection{Computational cost}

The computational cost at each iteration of the MCMC is dominated by evaluating the likelihood.
Depending on the model configuration, TraitLab uses one of two computational schemes:
\begin{itemize}
\item In the absence of lateral trait transfer ($ \beta = 0 $), \citet{nicholls2008} describe how to efficiently compute the non-zero terms in \eqref{eq:int-like} directly using a variant of Felsenstein's pruning algorithm \citep{felsenstein1981}.
The computational cost of this likelihood calculation is linear in the number of taxa and the number of site patterns.
Under the model with no lateral transfer, TraitLab's implementation thus allows for a large number of taxa (at least several hundreds).
\item When lateral trait transfer is included in the model ($ \beta > 0 $), the likelihood has the same form as in \eqref{eq:int-like}. However we cannot evaluate the expected pattern frequencies $ (z_p : p\in \cP) $ through recursions so must instead solve an expanding sequence of initial value problems \citep{kelly2017lateral}. In the final linear system, there is one differential equation for each pattern in $ \cP $.
As $ \abs{\cP} = 2^L - 1 $, the computational cost of evaluating the likelihood to machine precision grows exponentially with $ L $.
Consequently, computation with the lateral transfer model is only feasible for $ L \leq 20 $ taxa.
\end{itemize}

The family of MCMC proposals in \tabref{tab:moves} are designed to be computationally cheap per MCMC iteration at the possible expense of sampling efficiency.
The lag-coupled MCMC implementation in TraitLab can be used to accurately assess how well the MCMC algorithm explores the posterior for a given problem.
Sampling from the coupled Markov proposal kernel for a pair of chains takes a little over twice as long as sampling from the marginal kernel for a single chain, and again the likelihood calculation dominates the computational cost since currently TraitLab computes it in serial for each chain.

\section{Package}

\subsection{Installing and running TraitLab}\label{sec:install}

\subsubsection{System requirements}

TraitLab is written in the Matlab programming language.
Matlab is proprietary software which runs on Windows, Mac OS X or Unix/Linux machines.
TraitLab only requires the basic Matlab installation and its Statistics and Machine Learning toolbox.

\subsubsection{Download and installation}

Download the TraitLab code directory as a ZIP file from \url{https://github.com/traitlab-mcmc/TraitLab} and extract its contents preserving the subdirectory structure.
At the top level of the TraitLab directory are functions to launch TraitLab and scripts to set up the environment.
Lower level functions are contained in sub-directories, the \texttt{README} files within describe their contents.

The borrowing likelihood calculation requires either the installation of the Matlab Communications toolbox or the \texttt{mex} compilation of substitute functions written in C in the \texttt{borrowing} directory.
This requires that Matlab's \texttt{mex} compiler has been setup.
The files may be compiled from the top level of the TraitLab directory by executing \texttt{mexFiles} in the Matlab command window.
The compiled functions have the same names and syntax as their Communication toolbox counterparts.

\subsubsection{Launching TraitLab}\label{sec:launching-traitlab}

Start Matlab at the top level of the TraitLab directory. The \texttt{startup.m} script is automatically executed to set the correct path.
Otherwise, start Matlab in the usual manner, make the TraitLab folder the current directory and execute \texttt{startup} in the Matlab command window.

To start TraitLab and bring up the main GUI, type \texttt{TraitLab} at the Matlab command line. \secref{sec:batch} describes how to start MCMC runs using an input file instead of the GUI, either through a command line or the Matlab command window, which is useful for running long chains or launching multiple experiments.

\subsection{Data format and loading data} \label{sec:data}

TraitLab accepts data in the Nexus file format which is standard in phylogenetic analysis; see
\citet{maddison1997} for a full description of the format.
The binary data matrix is recorded in the \texttt{Data} block, while any clade constraints are recorded in the \texttt{Clades} block which is specific to TraitLab.
Block names, formatting and other properties in the Nexus file are case insensitive in TraitLab; taxa labels are case sensitive.
In the following, we describe the structure of these blocks and provide an example of a data file in \secref{sec:dataex}.

\subsubsection{Data block}

The first command in the \texttt{Data} block must be the \texttt{Dimensions} command with values for \texttt{ntax} (the number of taxa $L$ above) and \texttt{nchar} (the number of characters or traits $N$ above) defined.

The \texttt{Format} command, where the missing character is defined, is optional.
If it is not present, the missing character is assumed to be `\texttt{?}'.
The gap character, `\texttt{-}' by default, may also be defined here but is not used in TraitLab.
All gaps will be reclassified as missing.

The \texttt{Matrix} command is compulsory.
Rows of the matrix are labelled with the taxa names and may not contain any whitespace characters.
The content of the matrix must be zeros, ones and question marks, to indicate absence, presence or missingness of the trait.
The matrix may be in standard format (where the rows are \texttt{nchar} characters long) or in interleaved format (where the matrix is split in sections of a manageable size).
If the matrix is interleaved, each section of the matrix must have the rows labelled by the taxa names and sections must be separated by blank lines.
Comments may only occur between interleaved sections of the matrix --- comments at the start or end of rows will cause errors.

The \texttt{Charlabels} command, followed by a list of exactly \texttt{nchar} trait names may also appear in the \texttt{Data} block.
All other commands in the \texttt{Data} block are ignored.

\subsubsection{Clades block} \label{sec:cladeblock}

\begin{figure}
    \includegraphics[width=0.7\textwidth]{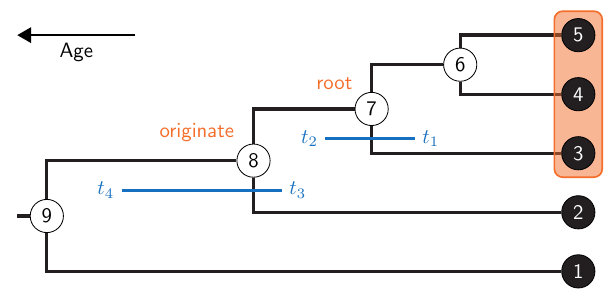}
    \caption{
        Suppose a clade comprises taxa 3, 4 and 5.
        If $ \texttt{rootmin} = t_1 $ and $ \texttt{rootmax} = t_2 $, then the age of the node labelled root, node 7 here, is constrained to lie in the interval $ [t_1, t_2] $.
        Similarly, if $\texttt{originatemin} = t_3$ and $\texttt{originatemax} = t_4$, then the age of the node labelled originate, node 8 here, is constrained to lie in the interval $ [t_3, t_4] $.
        Note that one child node of the originate node is necessarily the root node of the clade in question, but the other child node is arbitrary, here it is leaf node 2.
    }
    \label{fig:clade}
\end{figure}

Prior knowledge about the structure of the tree and divergence times can be encoded in the \texttt{clades} block.
The \texttt{clades} block consists of a series of clade commands, one for each known clade, and has the following form:

\begin{verbatim}
BEGIN Clades;

Clade Name = Clade_1
Taxa = taxon_a ... taxon_k
Rootmin = t1
Rootmax = t2
Originatemin = t3
Originatemax = t4;

Clade Name = Clade_2
Taxa = ...

End;
\end{verbatim}

Each \texttt{clade} command must include a name and the list of taxa that define the clade.
TraitLab will then force those taxa to be monophyletic: there must be a subtree consisting of exactly those taxa.
If desired, time constraints can be added to the clade.
The time of the most recent common ancestor (the root) of the clade can be bounded by defining \texttt{rootmin} and \texttt{rootmax}.
The time that clade diverged from all other taxa (the node above the clade root node) can be bounded by defining \texttt{originatemin} and \texttt{originatemax}.
See \figref{fig:clade} for an example of the difference between root and originate bounds.
All time bounds on a clade are optional.

\paragraph{Offset leaves}

If the taxa are sampled at significantly different times, then the \texttt{clades} block is used to encode this information.
If a clade has only one taxon, then the \texttt{rootmax} and \texttt{rootmin} definitions for that clade are upper and lower bounds for the sampling time of that taxon.
Time zero is defined as the time of the most recently sampled taxon.
The upper and lower time bounds must not be the same, so if a taxon was known to have been sampled 500 years before the most recently sampled taxon, set \verb|rootmax = 501| and \verb|rootmin = 499|.

\subsubsection{Other blocks}

When data are synthesized using TraitLab, a \texttt{trees} block and a
\texttt{synthesize} block are generated.
The \texttt{trees} block contains the tree on which the data were synthesized and the \texttt{synthesize} block contains the parameter values used for the synthesis.
If either of these blocks is found, TraitLab will assume that the data are synthetic.

The \texttt{characters} block may contain taxa names and/or trait labels.
Although it can be read by TraitLab, we advise against its use.
Include any relevant information in the \texttt{data} block instead.
All other blocks are ignored by TraitLab.

\subsubsection{Example data file}\label{sec:dataex}

The simple data file shown below shows the basic structure of a TraitLab data file.
It has the required \texttt{data} block and a \texttt{clades} block.
The \texttt{data} block specifies that there are 9 taxa and 30 traits then gives the data matrix.
The \texttt{clades} block specifies two clades, one with two taxa and with maximum and minimum age bounds on the root, the other with three taxa and a lower bound on the time it split from the rest of the tree.

\begin{verbatim}
#NEXUS

BEGIN DATA;

DIMENSIONS NTAX=9 NCHAR=30;
FORMAT MISSING=? GAP=-  INTERLEAVE ;

MATRIX

taxon_1  00?1111010110101000101001?0000
taxon_2  101111010?11001111001011011000
taxon_3  01101101??1110111?001010011000
taxon_4  010111100111011100110000100100
taxon_5  110111101111011??0110100100100
taxon_6  111111010111?01111001011011011
taxon_7  1111???101101011110??011011011
taxon_8  111111000111101110001011011011
taxon_9  11111101?111101111001011?11011
;
END;

BEGIN CLADES;

CLADE  NAME = Clade_1
ROOTMIN = 346
ROOTMAX = 422
TAXA = taxon_4, taxon_5;

CLADE  NAME = Clade_2
ORIGINATEMIN = 346
TAXA = taxon_1, taxon_8, taxon_9;

END;
\end{verbatim}

\subsection{Initializing an MCMC run} \label{sec:runmcmc}

\begin{figure}
    \centering
    \includegraphics[width=\textwidth]{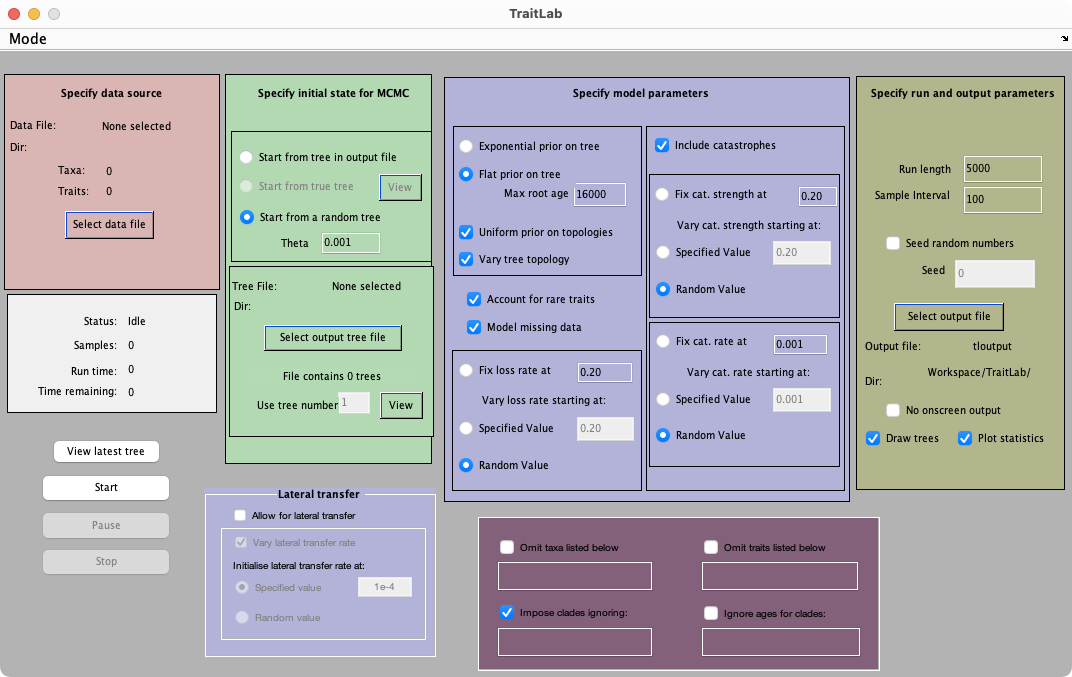}
    \caption{The main TraitLab GUI.}
    \label{fig:gui-main}
\end{figure}

MCMC analyses are set up and run in TraitLab from the main GUI shown in \figref{fig:gui-main}.
There are five types of information that the user must specify before starting a run: the data file, the initial state of the chain (which may be random), the model (including which parameters to estimate and priors), which parts of the data to ignore (if any), and the run parameters including run length and output file location.
These five types of information correspond to the five colours of the panels in the GUI and are each explained in detail in this subsection.

\subsubsection{Data source}\label{sec:datasource}

Specify the data source by clicking the \texttt{Select data file} button and choose a Nexus file conforming to the TraitLab requirements described in \secref{sec:data}.
Progress in reading the file can be monitored in the Matlab command window.

If the file is successfully read, the number of taxa and traits will be shown above the \texttt{Select data file} button.
These numbers should agree with the respective numbers of rows and columns in the matrix in the data file.
If any clades were found in the file, the number found is shown.
If the data were synthesized using TraitLab, this will also be indicated.

The names of the taxa and any clades found in the data file will appear in the Matlab command window.
The numbers that appear next to the trait and clade names are the order in which they appear in the file and are used to specify taxa and clades to be ignored.

\subsubsection{Omitting taxa and traits}\label{sec:omittaxa}

Any of the taxa and traits can be omitted from a run so long as at least two taxa remain to be analysed.
Omit taxa by checking the \texttt{Omit taxa listed below} checkbox and listing taxa numbers in the space provided.
A taxon's number corresponds to its row number in the data matrix in the data file.
Taxa numbers are also printed out to the Matlab command window after the data file is read in.
Omitting taxa that appear in clade constraints can cause problems; see \secref{sec:imposeclade} for more details.

Similarly, traits are omitted by checking the appropriate check box and listing a vector of trait numbers.
Trait numbers correspond to columns of the data matrix as they appear in the data file.
It is left to the user to find the correct column numbers for traits to be masked.

The lists of numbers must satisfy Matlab formatting; that is, numbers must be separated with commas or spaces.
Sequences of consecutive numbers can be abbreviated using the colon operator (for example, abbreviate \texttt{2 3 4 5} as \texttt{2:5}).
Thus, to omit taxa 1, 10, 11, 12, 13 and 20, type \texttt{1 10:13 20} in the space provided.

\subsubsection{Initial tree}

The initial tree may be sampled at random from the prior, chosen from a previous TraitLab run or, if the data were synthesized within TraitLab, the true tree on which the data were synthesized.

If the \texttt{Start from a random tree} option is selected, an Exponential tree (that is, a tree with exponentially distributed times between branching events) with specified branching rate $ \theta $ (so that the mean branch length is proportional to $ 1 / \theta $) is generated, then a short MCMC chain targeting the prior is run to create the initial state for the main analysis.
If clades are imposed, then the random tree is constructed using a heuristic method to ensure that all constraints are satisfied.

To start from a tree that appeared in a previous run, the output file where the tree is to be found needs to be specified using the \texttt{Select output tree file} button.
Once the file has been selected, a tree in the file is specified by the \texttt{Use tree number} text box.
Use the adjacent \texttt{View} button to view the specified tree.
The trees in the file must have taxa names which are a superset of the taxa names in the current run.

Note that if the initial tree is chosen from an output file, the user should ensure that the chosen tree satisfies all constraints to be imposed in the new run including the maximum root age and any clade constraints.
Specifying a tree that does not satisfy all the necessary constraints will result in a failed run.

If a run is started from a state in an output file which did not have lateral transfer in the model but the new run does, then TraitLab will return an error as there is no valid initial value for $ \beta $ in the initialising output file.
Only the number of catastrophes per branch is written to file and not their locations along branches.
Thus, when starting a run with lateral transfer from an output state, the location of each initial catastrophe is sampled uniformly along its corresponding branch.

\subsubsection{Model specification}\label{sec:modelspec}

The options in this section allow the user to decide how the data are to be modelled.
The following can all be included in or excluded from the fitted model: the catastrophe process, the modelling of missing data and the observation process that accounts for the fact that rare traits found in only a single taxon may not be observed.
In addition, the user must specify here which of the model parameters, $ g $, $ \mu $, $ \kappa $ and $ \xi $ are to be estimated.
The catastrophe rate $ \rho $ can be integrated out analytically or fixed at a specified value.

Radio buttons are used to determine the prior imposed on the tree.
We recommend the default prior which induces a uniform marginal prior distribution on the root age as described in \eqref{eq:unif_tree_prior}.
If the user selects this option, an upper bound on the root age $T$ must be specified in the \texttt{Max root age} box.
If clades are imposed, the specified value of $T$ must be greater than the largest of the \texttt{rootmin} and \texttt{originatemin} bounds imposed. One may additionally weight this prior so that its distribution over topologies is uniform.
The other tree-prior option is the so called ``exponential prior'' in which trees are given prior weight according to branch lengths which have an exponential penalty, as in \eqref{eq:exponential_tree_prior}.
If the \texttt{Vary tree topology} box is unchecked, the tree will remain in the same shape throughout the run with only the node ages varying.

The \texttt{Account for rare traits} checkbox determines whether or not the likelihood takes into account an observation process in which traits that only occur in a single lineage are recorded.
Check the box in the case in which the data collection process discards traits observed at only one taxon, as described in \secref{sec:obs}.
In this case, if the data matrix contains any traits that are observed in just a single taxon, those columns of the matrix are automatically removed before the analysis begins.
When the box is unchecked, the likelihood is calculated as if all traits observed in at least one taxon have been recorded.
Note that, for real data sets, it is not unusual for traits present in only one taxon to be discarded; if in doubt, we recommend that the checkbox remain checked.

If the data matrix includes traits coded as missing (that is, it includes \texttt{?} characters), the \texttt{Model missing data} box should be checked.
If it is left unchecked, then all missing traits will be recoded as absent and $ \xi $ parameters in the likelihood will be fixed at zero.

The trait death rate, $ \mu $, defined in \secref{sec:model} is an estimable parameter when clades with time constraints are imposed.
Instead of working directly with $\mu$, we reparametrize it as a \emph{loss rate} $ \psi $ defined by
\[
    \psi = 1 - \exp(-1000 \mu).
\]
Thus, $\psi$ is a number between $ 0 $ and $ 1 $ and can be interpreted as the mean proportion of traits that are lost in a lineage over a period of 1000 years.
The user has the option of fixing $\psi$, and thus $ \mu $, at a specified value or letting $ \mu $ vary over the run in order to estimate it.

If the data are thought to include rate heterogeneity, the catastrophe process may be included in the model by checking the \texttt{Include catastrophes} checkbox.
In this case, there are two further parameters to consider, the rate at which catastrophes occur, $\rho$, and the probability of trait death at a catastrophe, $\kappa$.
As with $\psi$, they can either be fixed at a specified value, or estimated during the run in the case of $ \kappa $ and integrated out in the case of $ \rho $.

The model allows for trait borrowing between branches at rate $ \beta $ when the \texttt{Allow for lateral transfer} checkbox is ticked.
Radio buttons and a text box allow the user to specify an initial value for $ \beta $ or let TraitLab initialise it with a uniform draw centred on the initial value for $ \mu $.
If the \texttt{Vary lateral transfer rate} checkbox is checked (the default), then $ \beta $ will vary over the course of the MCMC run, unchecking it means that $ \beta $ will be fixed at its initial value.


\subsubsection{Imposing clades}\label{sec:imposeclade}

If clade constraints are present in the data file, they may be applied during a run by checking the \texttt{Impose clades ignoring:} checkbox.

To omit clades from the analysis, list the clades to be omitted in the space provided using the respective clade numbers (see \secref{sec:omittaxa} for details on clade numbers and list formatting).
It is also possible to ignore the age ranges of certain clades (as if \verb|rootmin = 1| and \verb|rootmax = inf|) by checking the box \texttt{Ignore ages for clades} and listing the clade numbers in the space provided.

Difficulties can arise when omitting taxa that appear in imposed clades.
TraitLab warns the user about such cases in the Matlab command window and gives the option to either keep the clade with the taxon/taxa removed, or to ignore the clade completely.
Note that the first option can lead to errors, such as inconsistencies with other clades.


If problems are encountered omitting taxa that occur in clade constraints, the cleanest solution is to create another data file containing the same data block but a clade block that has been altered to achieve the desired constraints with the offending taxa removed.
This is the recommended course of action when the run is started in batch mode (see \secref{sec:batch}) as it is the only way to avoid user interaction.

\subsubsection{Run and output parameters}\label{sec:runparams}

The last panel that needs to be completed relates to the length of the run, the location of the output files and the required level of interactive monitoring of the run.

Specify the total length of the MCMC run, $r$, and sub-sample interval, $j$.
The initial state of the chain will be saved to the output files as will every $j$th state thereafter.
Thus the total number of sampled states that are saved is the integer part of $ (r / j) + 1 $ samples.
Choose $ r $ and $ j $ so that the total number of saved sampled states is somewhere in the range 1000--10000.
Sub-sampling reduces the correlation between samples and keeps the output files of a manageable size.

We are unable to give a priori guidance as to how long a particular run needs to be as it depends heavily on the particular data set in question.
Generally, the greater the number of taxa, the longer the run needs to be.
In some cases, it may take many days or weeks to obtain a satisfactory number of samples for larger problems.
Exploratory runs should be made to check convergence for the parameters of interest and it is sensible to make at least one very long run to check for any unexpected mixing behaviour in the chain.
Simple checks for a lack of convergence are discussed in \secref{sec:output}.
If the computational budget allows, then the coupling approach in \secref{sec:run-coupling} is a state-of-the-art approach for diagnosing convergence across all components of the tree and model.

The \texttt{Seed random numbers} checkbox is mainly used for debugging purposes and is generally left unchecked.
If checked, then a seed for the random number generator needs to be specified (it can be any real number) and will be recorded in the output \texttt{.par} file.
Separate runs with the same options and random seed will be identical.
If left unchecked, then the random number generator will be shuffled and no seed will be recorded.

\subsubsection{Output files}

Output files are saved in the location specified using the \texttt{Select output file} button.
If the default \texttt{tloutput} is the chosen file name, the output files created: \texttt{tloutput.txt}, \texttt{tloutput.par}, \texttt{tloutput.nex} and \texttt{tloutputcat.nex}.
If missing data is included in the model, then \texttt{tloutputXI.txt} is also created.
The coupling algorithm creates additional files which are described in \secref{sec:run-coupling}.

A record of all parameter values and options used to create the run are written to \texttt{tloutput.par}.
It can be used to perform similar runs in batch mode; see \secref{sec:batch} for details.

Each sampled tree is recorded in \texttt{tloutput.nex} and the corresponding parameter values \texttt{tloutput.txt} files.
The number of catastrophes on each branch of the trees in \texttt{tloutput.nex} are recorded in \texttt{tloutputcat.nex}.
Parameters recorded in \texttt{tloutput.txt} include the root time for the sampled tree, the trait death rate $\mu$, the death probability at a catastrophe $\kappa$ and the borrowing rate $ \beta $.
Note that since $ \lambda $ and $ \rho $ are integrated analytically, we return direct samples from their respective marginal posterior distributions at each iteration of the MCMC.
Also in the \texttt{.txt} file are unnormalized values of the log-prior density for the state, the integrated log-likelihood \eqref{eq:int-like} (where $\lambda$ has been integrated out) and the Poisson log-likelihood (using the sampled value of $\lambda$).
Finally, the $\xi$ parameters related to missing data are stored in the \texttt{tloutputXI.txt} file.

\subsubsection{Monitoring a run}\label{sec:monitor}

The progress of the run can be monitored via values output to the
Matlab command window, a plot of the most recently sampled tree and trace plots of the parameter and log density values.
There is also a status box on the left of the GUI showing the number of samples already obtained, the run time to date and an estimate of the time remaining.
See \figref{fig:semitic-initial-run} for an example.

To plot each tree as it is sampled, check the \texttt{Draw trees} checkbox.
If the box is unchecked, then the most recently sampled tree can be plotted by clicking the \texttt{View latest tree} button.

To see the trace plots of the sampled root time, trait loss rate,
log-prior density and integrated log-likelihood, check the \texttt{Plot statistics} checkbox.
Note that the log-likelihood trace plot is truncated to omit the early sampled values so that the vertical scale of the plot is reasonable.
If the data were synthesized using TraitLab, horizontal lines showing the true parameter values for the data are shown on the trace plots.

Unless the \texttt{No onscreen output} checkbox is checked, when each sample is taken, a row of numbers will appear in the Matlab command window.
These numbers show the sample number, the log-likelihood value and several statistics showing the proportion of proposed MCMC updates accepted since the previous sample.
For example, the line
\begin{verbatim}
(5,-3550.486804) 0.33 0.24 0.09 0.14 0.03 0.21 0.38 . . .
\end{verbatim}
means that sample 5 has an unnormalized log-likelihood \eqref{eq:int-like} of approximately $ -3550 $ and, of the updates between sample 4 and sample 5, 33\% of the states proposed by move type 1 were accepted, 24\% of those proposed by move type 2 were accepted and so on.
The indices of the moves are given in \tabref{tab:moves}.
Note that, depending on the options chosen, some moves may not be proposed so the proportion accepted will be \texttt{NaN}.

\subsection{Running TraitLab in batch mode}\label{sec:batch}

TraitLab can be run without the GUI interface from the Matlab command window or directly from a Unix, Linux or Mac command line.
Instead of using the GUI to specify all parameters for a run, they are read from a \texttt{.par} file.
Rather than write an input file from scratch, it may be more straightforward to set up a short model run using the GUI and then edit the resulting \texttt{.par} for the other runs in batch mode.

To run TraitLab without using the GUI, we use the function \texttt{batchTraitLab} whose first argument is the path to the \texttt{.par} file specifying the run parameters.
An optional second argument takes a numeric variable which is appended to the names of created files, so is useful when replicating experiments.

To start a run in batch mode from a command shell, make the main TraitLab directory the current working directory and execute
\begin{verbatim}
matlab -batch 'batchTraitLab("<path to .par file>")' > out.log
\end{verbatim}
at the command line, where the path to the \texttt{.par} file is a enclosed in double quotation marks and the entire command in single quotation marks.
Any onscreen output will be dumped to \texttt{out.log} while the normal \texttt{.nex}, \texttt{.txt}, \texttt{.par} and \texttt{XI.txt} output files will be created in the location specified in the input \texttt{.par} file.

TraitLab's batch mode is more restrictive than the GUI in initialising the model with catastrophes. If \verb|Random_initial_cat_death_prob = 1|, then the chain is initialized with $ \kappa \sim Unif(0.25, 1) $ and allowed to vary in the MCMC; otherwise $ \kappa $ is fixed at the value of \verb|Initial_cat_death_prob|.
Similarly, if \verb|Random_initial_cat_rate = 1|, then $ \rho $ is integrated out analytically; otherwise it is fixed at the value of \verb|Initial_cat_rate|.

As with the GUI, when initialising a chain with a sample from a previous run, the MCMC will ignore any specified initial value for a parameter in the input \texttt{.par} file, but will inherit whether the parameter is fixed or allowed to vary.

\subsection{Coupled MCMC algorithm} \label{sec:run-coupling}

In order to run a pair of lag-coupled Markov chains, edit the input \texttt{.par} file to set \verb|Coupled_markov_chains = 1| and \verb|Coupling_lag = <lag>|, where \verb|<lag>| is an appropriate choice of lag $ l $.
It is a requirement of TraitLab that $ l $ be an integer multiple of \verb|Sample_interval|.
The chains will run for \verb|Run_length| iterations or until the chains meet at random time $ \tau^{(l)} $, whichever is greater.
In the event that $ \tau^{(l)} < \verb|Run_length| $, TraitLab will only run the $ X $ chain for the iterations after $ \tau^{(l)} $ since the $ Y $ chain samples are identical to $ X $ after meeting.
As TraitLab only checks whether the chains have met when storing samples according to the sub-sample rate, the recorded value of $ \tau^{(l)} $ is slightly conservative.

We then run a pair of coupled chains via \texttt{batchTraitLab}.
As we typically run many pairs of coupled chains, we use the second argument of \texttt{batchTraitLab} to add a suffix to the corresponding output files.
Once the algorithm starts drawing coupled samples, TraitLab alternates between printing the sample index and log-likelihood for the $ X $ chain and the $ Y $ chain.
The coupled analyses produce a similar set of output files to the marginal experiments, except \texttt{\_x} or \texttt{\_y} is appended to the corresponding filenames.
A \texttt{.tau} file records the meeting time divided by the sub-sample rate, so we multiply the value in the \texttt{.tau} file by the sub-sample rate to obtain $ \tau^{(l)} $.

Increasing the lag $ l $ reduces the variance of $ \tau^{(l)} $ and thus provides better estimates of the total variation bound.
As the bound is not tight, there is a diminishing return of increasing the lag; \citet{biswas19} advocate increasing the lag until estimated total variation bounds stabilise.
We recommend running a sufficient number of coupled pairs of chains for each lag to obtain the desired level of confidence in estimates.
The \texttt{example} directory contains a synthetic data set \texttt{L8-data.nex} and parameter files for coupled and marginal experiments, \texttt{L8-coupled.par} and \texttt{L8-marginal.par} respectively.
Further guidance on running coupled experiments is provided by \citet{kelly2023lagged}.
Scripts to set up and run coupled experiments, analyse their output and estimate convergence bounds are available at \url{https://github.com/lukejkelly/CoupledPhylogenetics}.

Moves which rescale multiple node times by a single random factor (moves 6, 7 and 12 in \tabref{tab:moves}) cannot be coupled efficiently so are disabled when running the coupled MCMC algorithm.
Similarly, the coupling performs better when constraints distinguish rate parameters from time parameters; for example, \citet{kelly2023lagged} fix the death rate $ \mu $ and only infer node ages in their synthetic data analyses.

\subsection{Analysing output and inspecting the data} \label{sec:output}

\begin{figure}
    \centering
    \includegraphics[width=\textwidth]{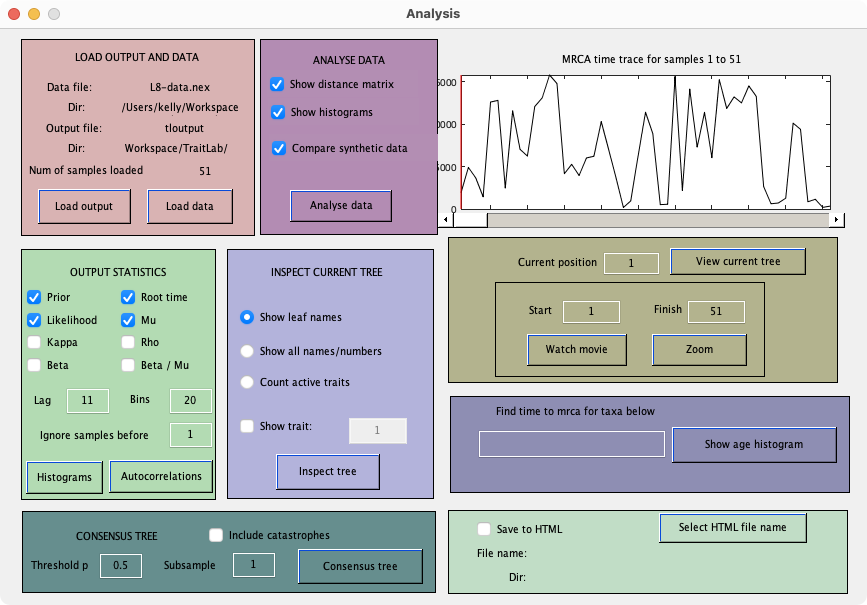}
    \caption{The analysis GUI.}
    \label{fig:gui-analysis}
\end{figure}

The results of MCMC runs and the raw data can be inspected in the analysis GUI, shown in \figref{fig:gui-analysis}.
To access the analysis GUI, choose \texttt{Analyse output} from the \texttt{Mode} menu of the main TraitLab GUI.

The output inspection tools described in \secref{sec:mcmcout} allow the user to explore the output of MCMC runs, including viewing sampled trees that are saved in a \texttt{.nex} output file, making trace plots and histograms of sampled statistics from a \texttt{.txt} output file and making histograms of the root time of user specified clades through the run.
These functions are described in \secref{sec:mcmcout}.
Note that many of these tools are fairly generic and can equally well be performed, along with other functions not available in TraitLab, by software such as Tracer \citep{rambaut18tracer}.

The data inspection tools described in \secref{sec:inspdata} allow the user to make histograms of the number of traits per taxon and of the number of taxa per trait, comparisons with synthetic data and plots based on a maximum a posteriori estimator of the time of the most recent common ancestor for a pair of taxa.
All of the plots made in TraitLab can be saved or printed as one would a standard Matlab figure.

\subsubsection{Loading data and output to the analysis GUI}

When the analysis GUI is called, the output from the most recent completed run (if there is one) along with the currently loaded data is automatically loaded into the analysis GUI.
If no run is in memory, nothing is passed to the analysis GUI and data and output can be loaded using the \texttt{Load output} and \texttt{Load data} buttons.
Note that it is left to the user to ensure that the loaded output and data are compatible.

\subsubsection{Exploring MCMC output}\label{sec:mcmcout}

\paragraph{Autocorrelations, trace plots and histograms}

The \texttt{Output statistics} box provides functions for calculating and plotting histograms, autocorrelation functions and trace plots of key parameters and statistics.
Visual inspection of the trace plot of a parameter of interest is a simple way to determine if a chain is still converging towards is target.
Similar plots can be drawn for the age of internal nodes; see below for a description.
For further information about how to interpret autocorrelation plots and associated statistics, see \citet{geyer1992}.
See \citet[Section 2.2]{kelly2023lagged} for an extensive overview of methods to diagnose MCMC convergence in phylogenetic problems.

%

To plot the autocorrelation functions, it is necessary to specify the maximum lag for which autocorrelations are calculated in the \texttt{Lag} text box.
To plot marginal histograms of the sampled statistics, specify the number of bins to use in the histogram.
A burn-in can be specified using the \texttt{Ignore samples before} text box.

\paragraph{Viewing sampled trees}\label{sec:viewtree}

At the top right of the GUI is a trace plot of the log-likelihood of sampled states in the loaded run.
Trees from the loaded run can be viewed using the \texttt{View current tree} button and (when data are also loaded) more closely investigated using the \texttt{Inspect current tree} functions.
The current tree is the tree at the current position, which is shown as a red line on the log-likelihood plot and can be specified either in the \texttt{Current position} text box or by using the slider below the trace plot.

When data are loaded, the \texttt{Inspect tree} function can be used to closely inspect sampled trees.
If the \texttt{Show leaf names} radio button is selected, the standard plot of the current tree is drawn.
The \texttt{Show all names/numbers} option is mainly used for debugging and shows the node numbering used to internally represent the tree.
The numbers on the leaf nodes shown in this representation are those used to specify taxa for the MRCA histograms.

The \texttt{Count active traits} option shows the standard tree plot of the current state where, at each internal node of the tree, the number of traits found exclusively in the clade defined by that node is shown.
For example, suppose that the current tree has three taxa, \emph{a}, \emph{b} and \emph{c}, of which \emph{a} and \emph{b} form a clade, and there are 20 traits found in \emph{a} and \emph{b} that are not found in \emph{c}.
Then the number 20 will be shown at the parent of \emph{a} and \emph{b}.
Note that the number shown at the root of the tree is always the total number of traits in the data set being analysed after rare traits have been stripped from the raw data.

By checking the \texttt{Show trait} checkbox and specifying a trait number, the path of that trait on the current tree can be viewed.
According to the standard Dollo model, the trait must have been present on the lineages shown in black, so must have been born either above the root of the tree or on the lineages shown in red.
It was either not present or died at some point on one of the lineages in blue.

\paragraph{Histogram of MRCA for given taxa}\label{sec:MRCAhist}

The marginal posterior distribution of the time to the most recent common ancestor (MRCA) for any subset of taxa in the analysis can be estimated using the \texttt{Show age histogram} button.
Specify the taxa of interest by using their respective numbers in the text box provided.
The taxa numbers can be found using the \texttt{Inspect tree} function with the \texttt{Show all names/numbers} option selected.
%

\subsubsection{Inspecting the data}\label{sec:inspdata}

Clicking the \texttt{Analyse data} button will produce a number of plots which are described below.
Note that both data and output need to be loaded to view these plots since a tree is required for the construction of the distance-depth plot.

\paragraph{Data histograms}

If the \texttt{Show histograms} checkbox is checked, histograms of the number of taxa per trait and the number of traits per taxon are generated.
The taxa per trait histogram is a histogram made from the column sums of the data matrix.
The traits per taxon histogram is made from the row sums of the data matrix.
It is the empirical distribution of the number of traits found in each taxon.

If the \texttt{Compare synthetic data} checkbox is checked, synthetic data according to the standard stochastic Dollo model are generated on the current tree and histograms of the synthetic data are displayed below those for the loaded data.
If the model fits the data well and the tree is a representative draw from the posterior distribution, then the two pairs of histograms should be qualitatively very similar.
On the contrary, consistent qualitative differences between the two pairs of histograms may indicate some significant model misspecification.

\paragraph{Distance depth and distance matrix plots}

\begin{figure}
    \centering
    \includegraphics[width=0.7\textwidth,trim=2cm 8.5cm 2cm 8cm,clip]{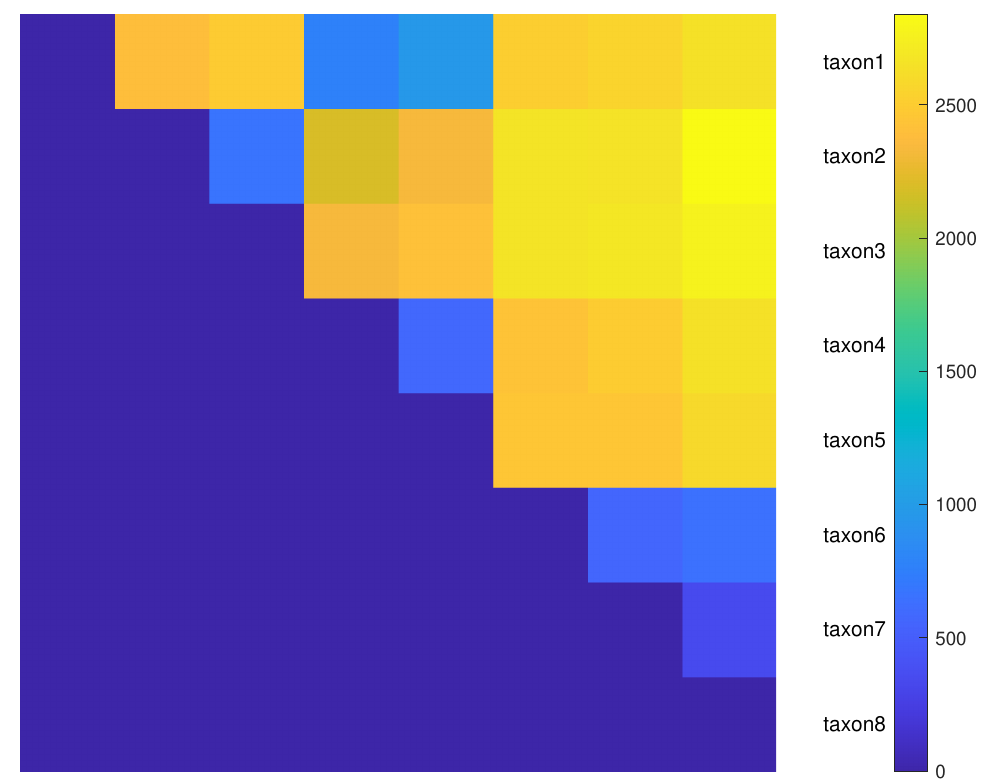}
    \caption{
        An example of the distance matrix graph.
        Row labels are the taxa names, the column labels are the same as the row labels.
        The pixel at row $i$, column $j$ represents the maximum a posteriori estimate of the MRCA between taxon $i$ and taxon $j$.
        The colour bar displays the corresponding distance for each colour.
        Note that the ordering of taxa is as in the data file, so the obvious clustering that is shown here will not occur if rows in the data file are randomly ordered.
    }
    \label{fig:distmat}
\end{figure}

\begin{figure}
    \centering
    \includegraphics[width=0.7\textwidth,trim=1cm 0.5cm 1cm 0.5cm]{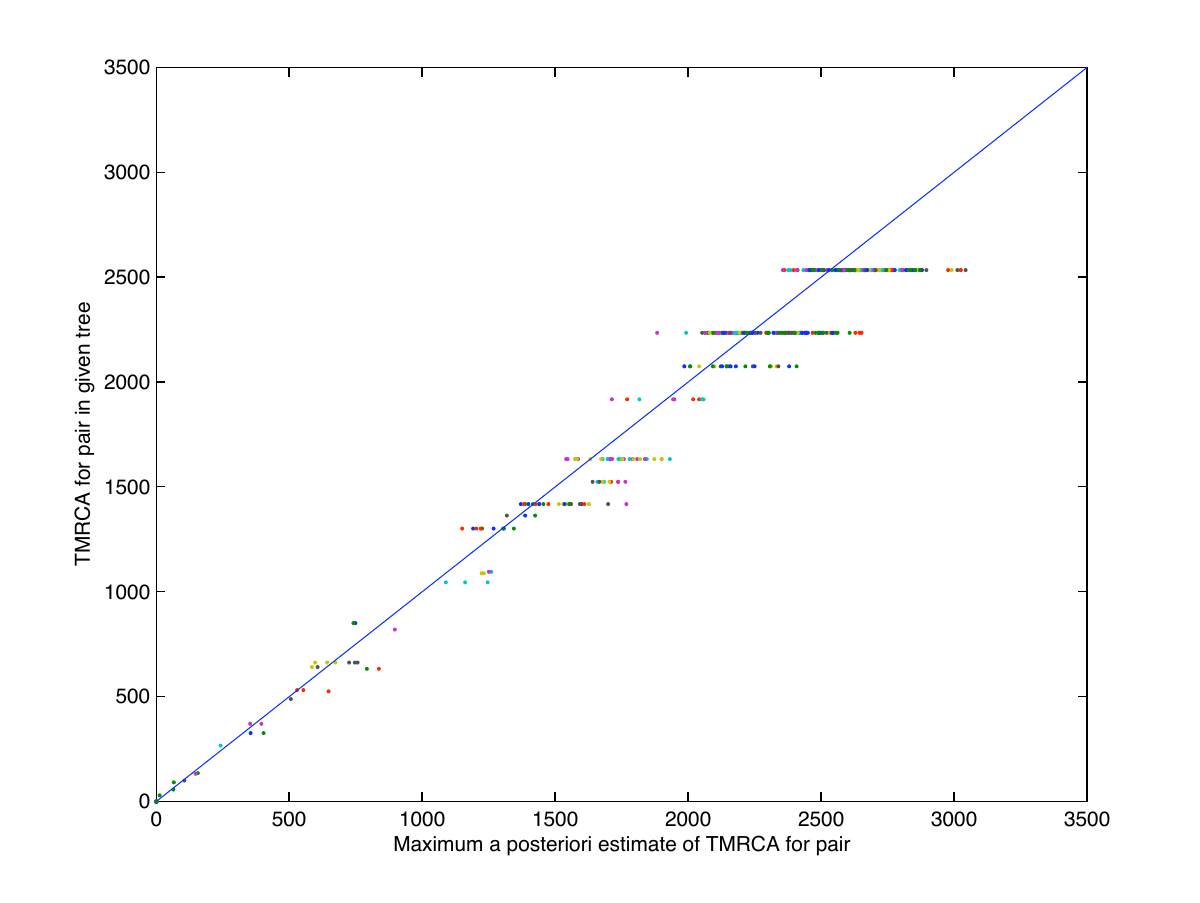}
    \caption{
        An example of the depth distance relation graph.
        Each point represents a pair of taxa.
        The position of the point along the $x$-axis is the maximum a posteriori estimate of time to the most common ancestor for the given taxa pair, while the position along the $y$-axis is given by the time of the most recent common ancestor of the pair in the given tree.
        The line shown in blue is $y=x$.
        The points are horizontally stratified as a single ancestral node in the tree may be the common ancestor of many pairs of taxa.
        The example here shows a very good fit of data and model.
    }
    \label{fig:depthdist}
\end{figure}

When all traits, including those found only in a single taxon, are observed it is possible to analytically calculate the maximum a posteriori estimate of the time to the most recent common ancestor for a pair of taxa under the standard Dollo model; see \citet[Equation~10]{nicholls2008} for further details.
The distance matrix plot in \figref{fig:distmat} and the distance depth relation plot in \figref{fig:depthdist} are based on this calculation.

The distance matrix graph is simply a heat plot of a matrix where this distance has been calculated for each pair.
It is best understood by referring to the example in \figref{fig:distmat}.

For the depth distance relation plot, the maximum a posteriori estimate for the MRCA for each pair is calculated and the time of the MRCA of each pair in the specified tree is found.
The two times for each pair are then plotted against each other.
See \figref{fig:depthdist} for an example.
The tree used is the current output tree (or, when the graph appears after synthesizing data, the tree on which the data were synthesized).
If the standard Dollo model and the specified tree fit the data well, then the points should lie along the line $y = x$.

\subsubsection{Building a consensus tree}

A consensus tree is a convenient way of summarizing a sample of trees.
A consensus tree shows all splits with at least 50\% support in the posterior sample.
If a split receives between 50\% and 95\% support, it is labelled; splits receiving at least 95\% support are unlabelled.
If no split receives more than 50\% support, then the tree is multifurcating.
The threshold can be changed to another value, but values lower than 50\% may lead to errors.
The parameter \texttt{Subsample} can be set to an integer value greater than 1 to subsample the output; this will speed up the computation of the consensus tree.

On a consensus tree, the length of a branch displayed is the average length of that branch in the trees of the posterior sample where it is present.
Similarly, the number of catastrophes displayed on a branch is the average number of catastrophes on that branch in the trees of the posterior sample where the branch is present, rounded to the nearest integer.

\subsubsection{Saving as HTML}

To ease the sharing of results, the GUI offers an option to save your analysis as an HTML file.
Tick the box \texttt{Save figures in HTML file}, then either select the name of your HTML file or let TraitLab choose one for you, based on the date and time.
So long as the check-box is selected, all figures you produce from the Analysis GUI are saved as \texttt{.bmp} files, and an HTML file is created with those figures and some brief explanatory text.
Alternatively, figures may be saved to other formats using the \texttt{print} functionality within Matlab.

\subsection{Synthesizing data and model checking} \label{sec:synth}

TraitLab allows the user to simulate trait and clade data under any of the fitted models, such as the Dollo model with or without catastrophes and with or without missing data, and also allows simulation under various model extensions.
The model extensions include borrowing (lateral transfer of traits between lineages), heterogeneity in the trait death rate, $\mu$, across edges in the tree and heterogeneity in $\mu$ across different groups of traits.
In this section, we provide  instructions on how to synthesize data in TraitLab, which involves specifying a tree, a model of trait evolution and a clade model.
We give full descriptions of each of the model extensions as we go along.

\begin{figure}
    \centering
    \includegraphics[width=\textwidth]{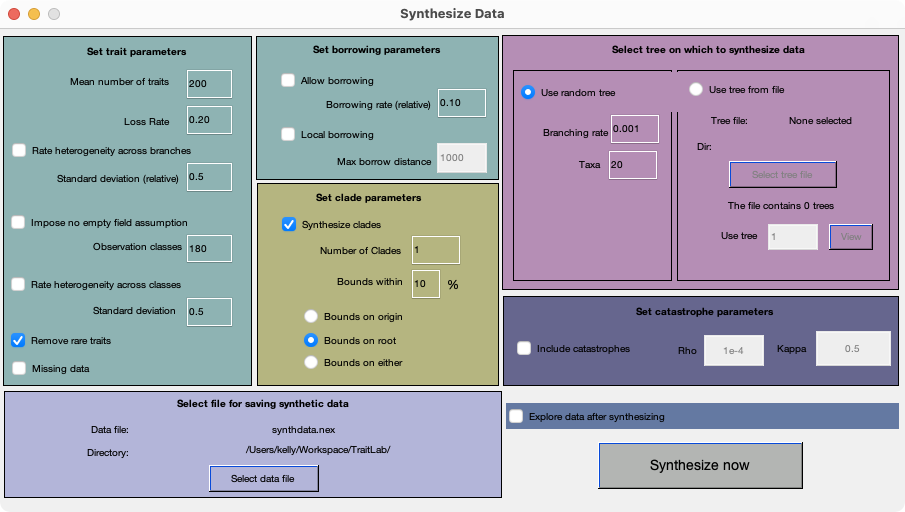}
    \caption{The data-synthesizing GUI.}
    \label{fig:gui-synthesize}
\end{figure}

To get started with the synthesize GUI, choose \texttt{Synthesize data} in the \texttt{Mode} menu from the main TraitLab GUI.
The synthesise GUI is shown in \figref{fig:gui-synthesize}.

\paragraph{Selecting a tree}

The tree along which traits evolve can either be a randomly generated exponential tree (with exponential branch lengths) or any rooted tree (with branch lengths specified) stored in the Newick format in a Nexus file.


\subsubsection{Defining the trait evolution model}\label{sec:synparams}

All models are based on the basic Dollo model of evolution, which is completely specified on a given tree by defining the trait birth and death rates.
Equivalently, we parameterize the model using the mean number of traits per taxon, $K$, and the trait loss rate, $\psi = 1 - \exp(-1000\mu)$.
Given the mean number of traits per taxon, $K$, and a value of $\mu$, the per taxon trait birth rate, $\lambda$, is defined by $\lambda = K\mu$.

\paragraph{Catastrophes}

Catastrophes are included in the simulation when the \texttt{Include catastrophes} checkbox is checked.
In this case, the catastrophe occurrence rate, $\rho > 0$, and the probability of trait death at a catastrophe, $ 0 < \kappa \leq 1 $, must be specified in their respective text boxes.
The trait birth rate at a catastrophe is $ \nu = \kappa \lambda / \mu $.

\paragraph{Rate heterogeneity across branches}

Another form of rate heterogeneity, different from catastrophes, is to let the trait death rate $\mu$ vary across each branch of the tree so that for branch $i$, there is a specific death rate $\mu_i \sim \Gamma\left(\text{mean} = \mu, \text{variance} = (\sigma \mu)^2\right)$.
It is necessary to specify the relative standard deviation parameter, $ \sigma $.

\paragraph{Observation classes: the no-empty-field assumption}

If the \texttt{Impose no empty field assumption} box is checked, then TraitLab will produce data as if traits are collected based on pre-specified observation classes for which it is assumed every observed taxon will have at least one trait in each class.
This is called the no-empty-field assumption as we assume that each observation class is occupied by at least one trait in each taxon at all times.
This is a natural assumption with lexical data where linguists set out to collect words from different languages associated with a list of meanings, and is discussed by \citet[Section~9.5]{nicholls2008}.
The meanings here are the observation classes, so the assumption is that every language will have a word for each basic meaning.
If this option is checked, it is necessary to specify the number of observation classes $n_{\text{obs}}$.
This option will have the greatest effect on the simulated data when $n_{\text{obs}}$ is close to the mean number of traits per taxon, $K$.
Note that it is natural to set $ n_{\text{obs}} \leq K $.

If the no-empty-field assumption is imposed, then heterogeneity in trait death rates across observation classes may be incorporated by checking the \texttt{Rate heterogeneity across classes} box and specifying a relative standard deviation, $ \varsigma $.
If the no-empty-field assumption is imposed and the \texttt{Missing data} box is checked, data will go missing in blocks: within an observation class and for a given taxon, either all traits are observed or all are missing.
The interested reader is referred to Section~4.2.2 of \citet{ryder2010dphil}.

When rates vary across both branches and observation classes, a trait on branch $i$ in observation class $j$ dies at rate $\mu_{ij} \sim \Gamma(\text{mean} = \mu_i, \text{variance} = (\varsigma \mu_i)^2) $, where $ \mu_i \sim \Gamma(\text{mean} = \mu, \text{variance} = (\sigma \mu)^2) $ as above.

\paragraph{Borrowing}

Borrowing, or horizontal transfer of traits, is observed in lexical, cultural and genetic trait data.
We allow users to simulate data under two models of borrowing --- global or local borrowing.
Global borrowing is when all lineages are equally likely to borrow from one another, whereas local borrowing means that only lineages that are close, in the sense that they share a recent common ancestor within some specified time period, may borrow traits from each other.

Global borrowing is synthesized by checking the \texttt{Allow borrowing} checkbox and specifying a relative borrowing rate, $b$.
Each trait is borrowed at constant rate $b \mu$.
When a trait is borrowed, it first chooses a target branch $i$ uniformly at random from the extant lineages. If the trait is not currently present in lineage $i$, then it is added to $ i $, otherwise nothing happens.

To make borrowing local instead of global, check the \texttt{Local borrowing} checkbox and specify a borrowing distance, $d$.
All traits are still borrowed at constant rate $b \mu$ but the target lineage is chosen uniformly at random from all lineages which share a common ancestral lineage with the source lineage in the last $d$ years.

When death rates are heterogeneous and borrowing occurs, a trait on branch $i$ in observation class $j$ is borrowed at rate $ b \mu_{ij} $.

\paragraph{Missing data and rare traits}

Once trait data have been simulated down a tree according to the specified model, the observation model is simulated so that some traits may be marked missing and rare traits are ignored.
If the \texttt{Include missing data} checkbox is checked, a parameter $ \xi_i \sim \beta(1, 1/3) $ is drawn for each leaf $i$.
At leaf $i$, each entry in the simulated matrix is then made missing with probability $1-\xi_i$.
Columns of the matrix with no 1s (traits not observed at any taxa) are removed.
If the \texttt{Remove rare traits} checkbox is checked, traits that are observed at only one taxon are also discarded.

\paragraph{Synthesizing clades}

Check the \texttt{Synthesize clades} checkbox to synthesize clades.
When there are $L$ taxa in the synthetic data, up to $ L - 1 $ clades can be synthesized corresponding to the $ L - 1 $ internal nodes of the tree.
The accuracy of the clade bounds is specified in the \texttt{Bounds within} text box.
An accuracy of $c$ means that if the chosen node has a time $t$ in the tree, then a lower bound of $ (1 - c / 100)t $ and an upper bound of $ (1 + c / 100)t $ will be created.

\subsubsection{Output of synthesize GUI}

The synthetic data are saved to the \texttt{.nex} file specified by the user, the default is \texttt{synthdata.nex}.
It contains the synthetic trait matrix, the tree on which the data were synthesized and associated model parameters.
A \texttt{.par} file, \texttt{synthdata.par}, is generated for record keeping purposes: all options and parameter values used to produce the data are recorded here.
It is not the same as the \texttt{.par} file for an MCMC run.

\section{Summary}

TraitLab implements a stochastic Dollo model for binary data and various extensions for rate heterogeneity via catastrophes and lateral trait transfer.
It is well suited to data for which homoplasy (parallel evolution) is implausible.
It offers tools to verify the convergence of the MCMC and the model fit, as well as to simulate from a variety of generalized models.

\begin{appendix}
\section{Analysis of Semitic lexical data}
\label{app:semitic}

As an example, we provide in this section the steps used to analyse a real data set.
The data concern core vocabulary of various Semitic languages and were collected and made public by \citet{kitchen2009bayesian}.
The data set is included in TraitLab in the Nexus file \texttt{semitic09.nex} in the \texttt{example} directory.
An analysis of these data with TraitLab first appeared in \citet{nicholls2011phylogenetic}.

\paragraph{Data file}

The data file is in Nexus format.
The \texttt{DATA} block includes 674 traits for 25 taxa.
Each taxon is a Semitic language (Gehez, Tigre, Amharic, \dots) and each trait is a meaning category; the meaning categories cover 96 meanings from the core vocabulary.
In this file, some data points are listed as gaps (\texttt{-} in the data file); these are treated as missing by TraitLab.

The data file also includes a \texttt{CLADES} block with seven clades constraints.
In this case, all but one clade comprises a single language and thus gives information on one of the leaves of the tree.
A subset of the clades block is:
\begin{verbatim}
CLADE  NAME = hebrew
ROOTMIN =2500 ROOTMAX =2700
ORIGINATEMIN = 3200 ORIGINATEMAX = 4200
TAXA = Hebrew ;
\end{verbatim}
meaning that the sample of Hebrew in the data corresponds to a language spoken between 2500 and 2700 years Before Present, and that we know that the branch leading to Hebrew split off from its closest cousins between 3200 and 4200 years Before Present.

\paragraph{Loading the data}

Launch Matlab with TraitLab the current directory according to the instructions in \secref{sec:launching-traitlab}, then type \texttt{TraitLab} in the command window to launch the TraitLab GUI.
In the \texttt{Specify data source} panel, click \texttt{Select data} and navigate to the \texttt{semitic09.nex} file in the \texttt{example} directory.

In the Matlab command window, we are given a list of clades and a list of languages, as well as a list of languages with more than 5\% missing data.
The TraitLab GUI reports that the data file contains 25 taxa and 674 traits.

\paragraph{Initial run}

For now, we can keep all the parameter values unchanged and click the \texttt{Start} button to initiate the MCMC run.
By default, TraitLab will use a run length of 5000 iterations subsampled to every 100 iterations, thus giving a sample of size 50.

\begin{figure}
    \includegraphics[width=0.7\textwidth, trim=2cm 8.5cm 2cm 8cm, clip=true]{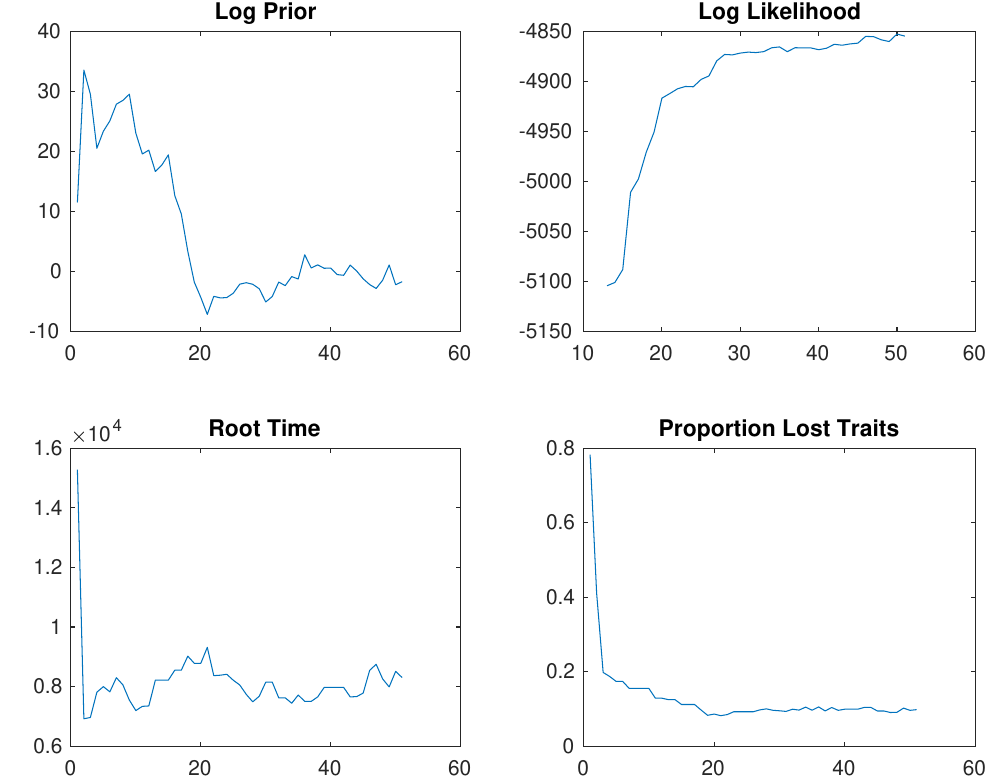}
    \caption{
        Trace of key statistics after the first MCMC run for the Semitic data.
        These plots (the log-likelihood, for example) show that the MCMC has not yet reached stationarity.
    }
    \label{fig:semitic-initial-run}
\end{figure}

Extra information is given in the Matlab command window.
First, the data are prepared.
By default, all traits which are absent at all taxa or present at only 1 taxon are removed from the data, leaving 334 traits.
During the run, the log-likelihood of each sample is displayed alongside the acceptance ratios for each move, and a figure of the current tree is plotted as well as traces of several key statistics, as shown in \figref{fig:semitic-initial-run}: the plot includes the traces of the log-prior, log-likelihood, root age, and proportion of lost traits (the proportion of traits which existed at some point on the tree but died before reaching any leaves).
It is immediately obvious that the MCMC has not converged for this short run, so we launch a longer run: five million iterations, subsampling to every 5000 iterations for a final sample size of 1000.

At first glance, this second run (not shown) seems satisfactory from the trace plots of key statistics.
We move to the \texttt{Analyse output} mode to further assess whether the chain has reached stationarity and mixed well.

\paragraph{Checking MCMC convergence}

\begin{figure}
    \includegraphics[width=\textwidth, trim=3cm 1.5cm 3cm 1.5cm, clip=true]{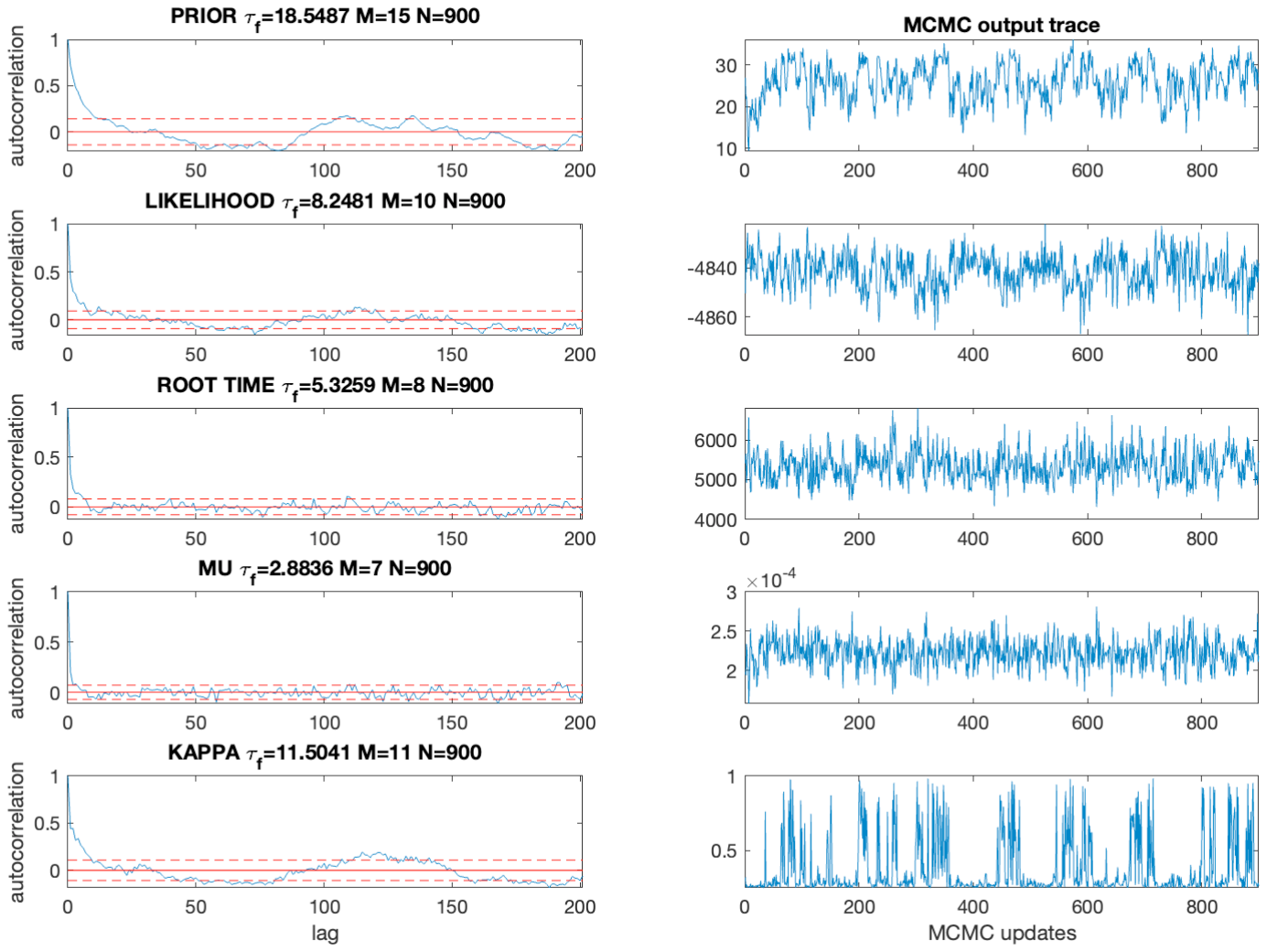}
    \caption{
        Autocorrelation and traces of parameters and statistics after a long MCMC run for the Semitic data.
    }
    \label{fig:semitic-long-trace}
\end{figure}

\begin{figure}
    \centering
    \includegraphics[width=0.7\textwidth, trim=3cm 1.5cm 3cm 1.5cm, clip=true]{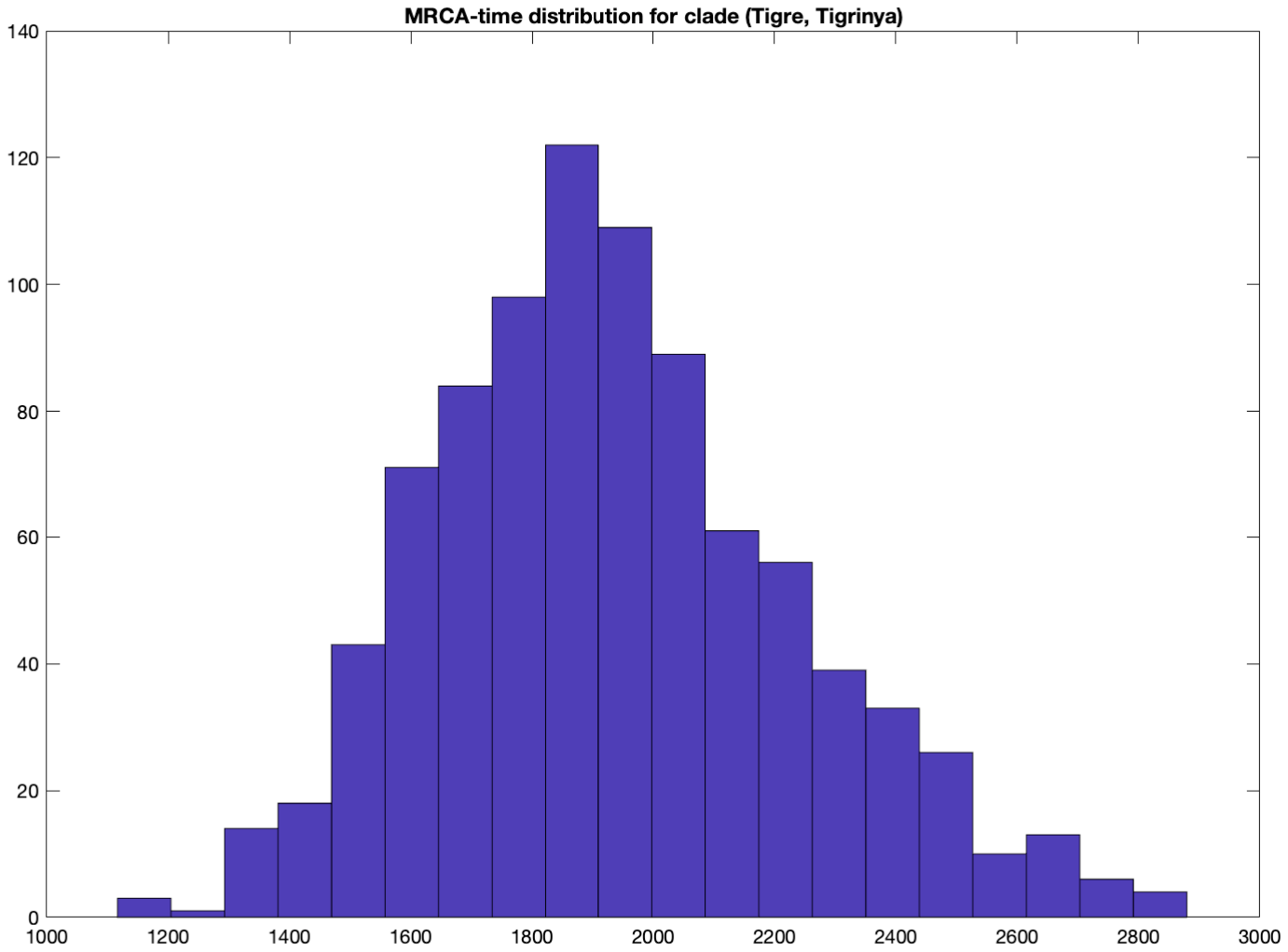}

    \includegraphics[width=0.7\textwidth, trim=3cm 1.5cm 3cm 1.5cm, clip=true]{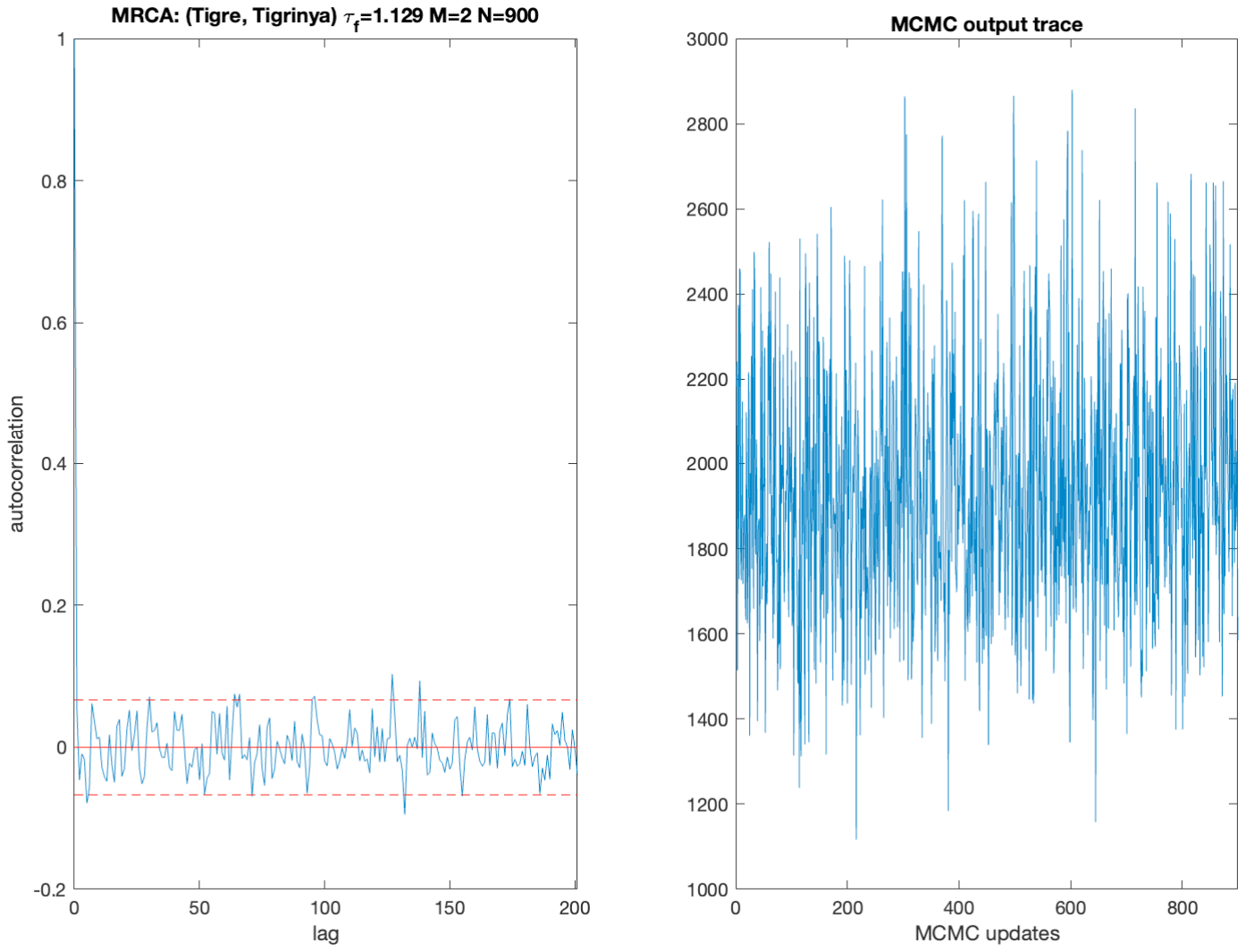}
    \caption{
        Histogram, autocorrelation and trace plots for the age of the MRCA of Tigre and Tigrinya after a long MCMC run for the Semitic data.
    }
    \label{fig:semitic-long-mrca}
\end{figure}

In the \texttt{Analyse output} mode, the data and run output are loaded automatically.
The log-likelihood is plotted over the entire run; we zoom in to display only iterations 102 to 1001.
The plot suggests that a burn-in of 100 iterations from the initial state is satisfactory.
In the green \texttt{Output statistics} panel, we check the trace and autocorrelation plots of the prior, root time and log-likelihood, as well as the parameters $\mu$ and $\kappa$, discarding the initial state and first 100 samples as burn-in.
These plots, displayed in \figref{fig:semitic-long-trace}, are also satisfactory.
Finally, we check the traces and autocorrelations of a few internal and leaf ages.
To do this, first use the \texttt{Inspect current tree} panel with the option \texttt{Show all names/numbers}.
A tree is displayed, with the numbers used by TraitLab to represent each leaf node internally.
These numbers can then be passed to \texttt{Show age histogram} to get the distribution of their MRCA.
For example, we check the age of the leaf Hebrew and the age of the internal node above Tigre and Tigrinya by entering the relevant node numbers.
Along with the age histogram, we are also given MCMC trace and autocorrelation plots for that node age the indicator that the languages given form a subtree (this gives an indication of the MCMC behaviour of the tree topology).
\figref{fig:semitic-long-mrca} displays these plots for the MRCA of Tigre and Tigrinya.
Summary statistics are also given in the Matlab command window; in this example, it reports an 85\% posterior probability that Tigre and Tigrinya form a clade.

These visual approaches may silently fail to diagnose a lack of convergence across all components of the model. Where the computational budget allows, we recommend using the couplings described in \secref{sec:coupling} and \citet{kelly2023lagged} to confidently diagnose convergence jointly across all components.

\paragraph{Checking model fit}

\begin{figure}
    \centering
    \includegraphics[width=0.7\textwidth, trim=2.75cm 1.5cm 3cm 1.75cm, clip=true]{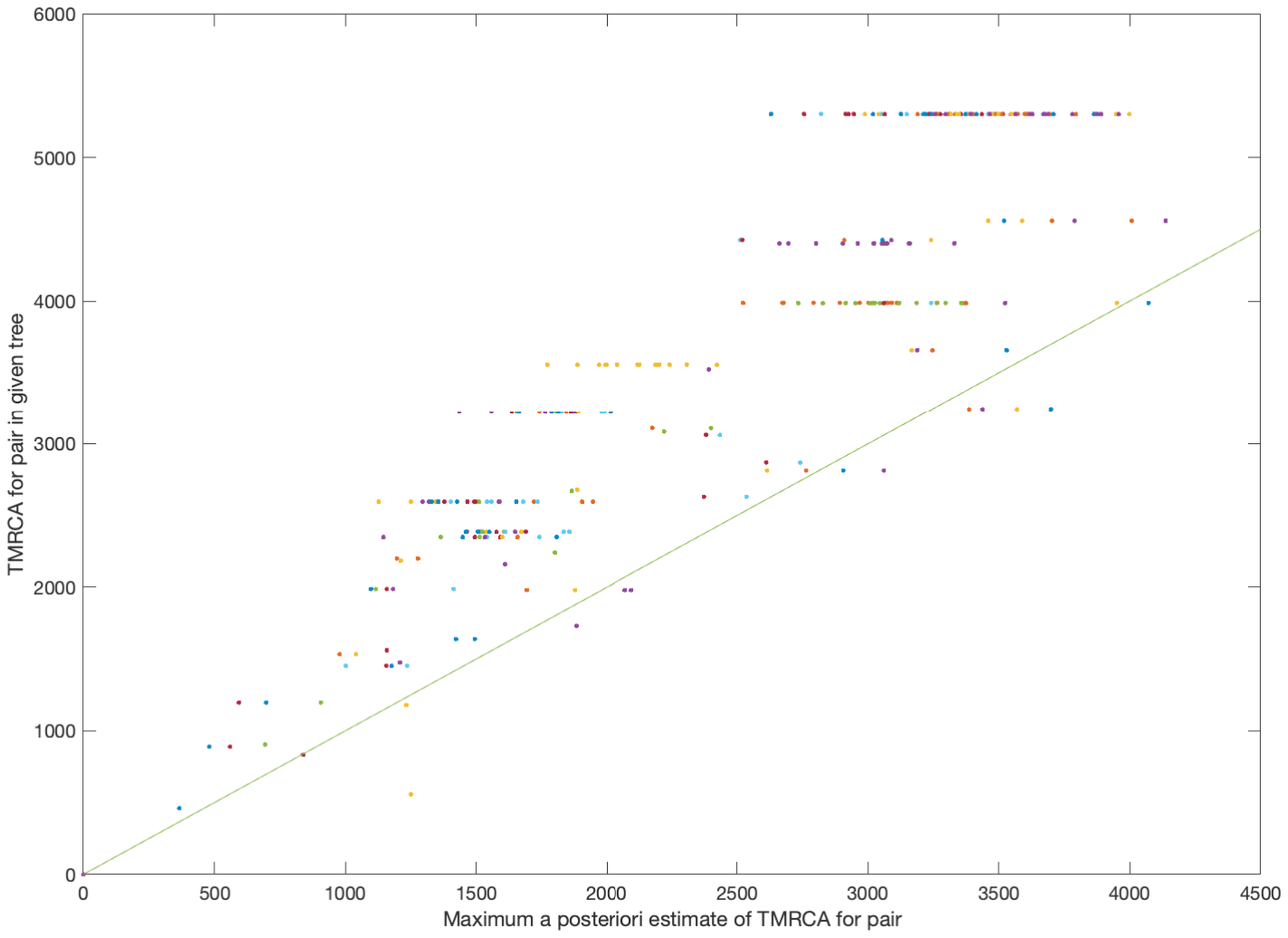}
    \caption{Distance depth relation plot for the Semitic data.}
    \label{fig:semitic-long-depth}
\end{figure}

\begin{figure}
    \centering
    \includegraphics[width=0.7\textwidth, trim=3cm 1.5cm 3cm 1.5cm, clip=true]{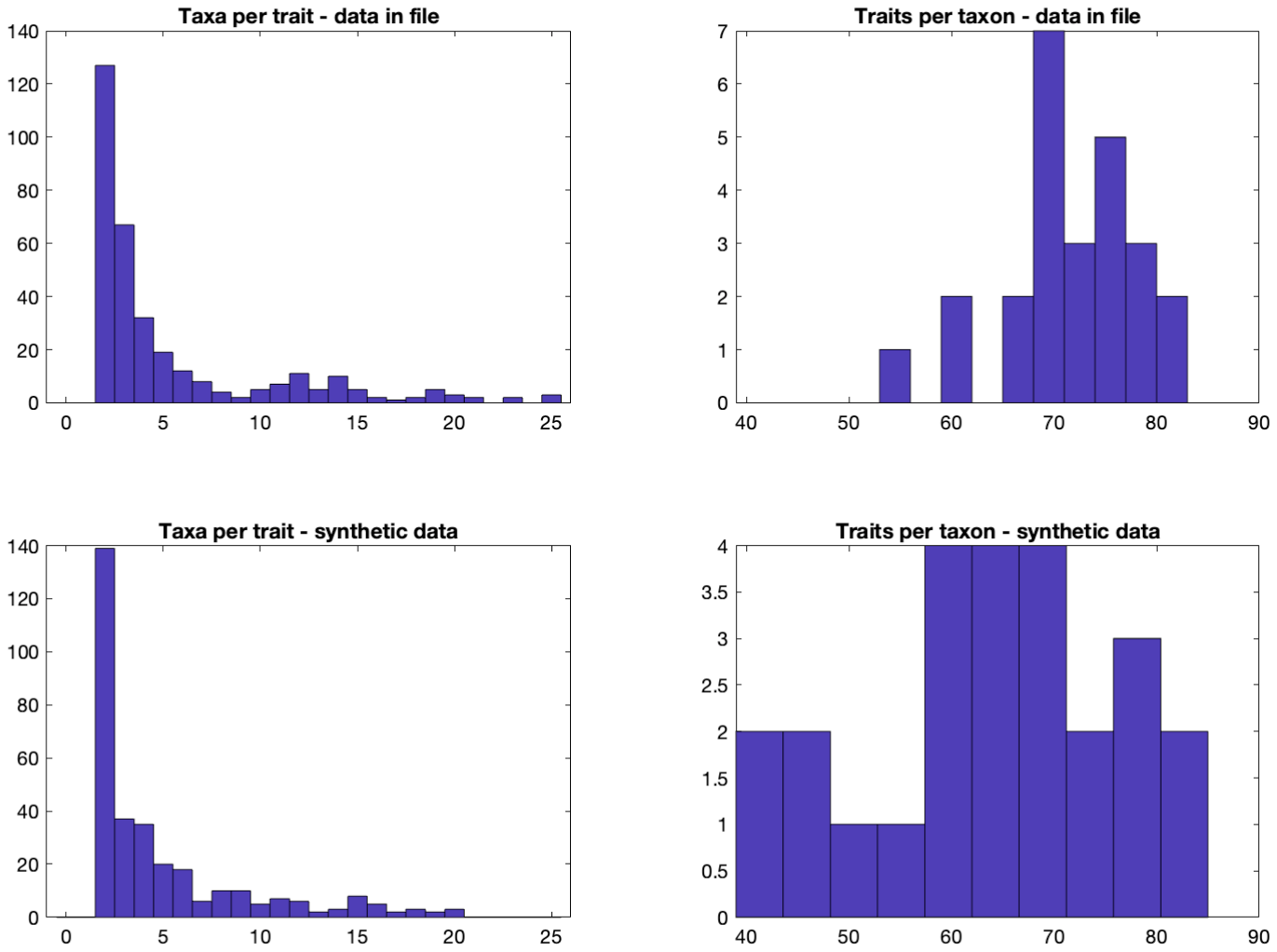}
    \caption{Comparing observed and synthetic trait frequencies for the Semitic analyses.}
    \label{fig:semitic-long-traits}
\end{figure}

Since the MCMC run appears to have converged, we can now check for model misfit.
Using the \texttt{Analyse data} panel, we observe in \figref{fig:semitic-long-depth} that points on the \texttt{Distance depth relation} plot, previously described in \figref{fig:depthdist}, are largely above the line $y=x$ but the histograms for the data in the file only partially resemble those for synthetic data \figref{fig:semitic-long-traits}, suggesting that the fit could be improved.

\begin{figure}
    \includegraphics[width=\textwidth, trim=0.5cm 0.5cm 3cm 0.5cm, clip=true]{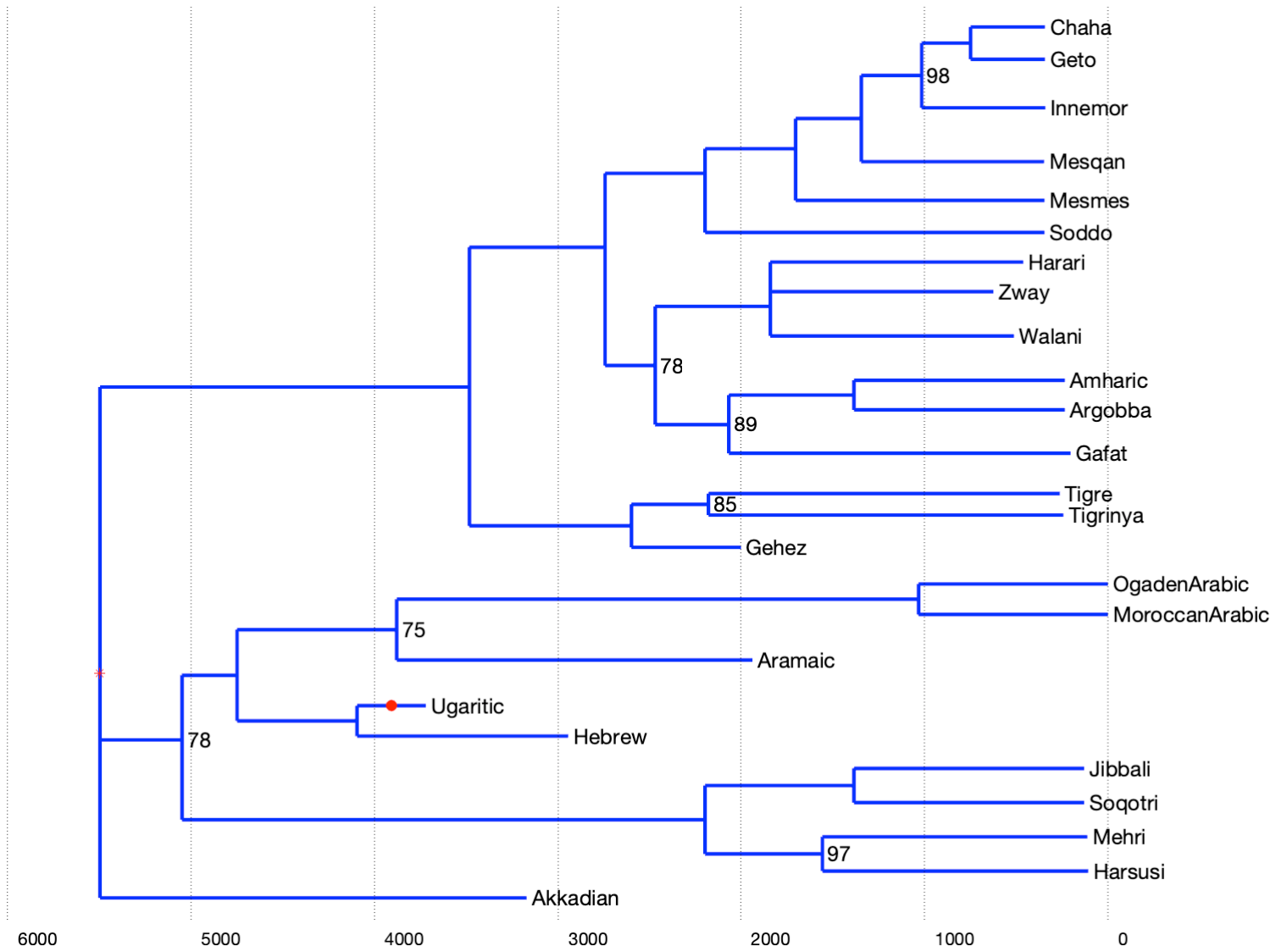}
    \caption{
        Consensus tree with a single catastrophe after a long MCMC run for the Semitic data.
    }
    \label{fig:semitic-long-consensus}
\end{figure}

The consensus tree with catastrophes in \figref{fig:semitic-long-consensus} shows strong support for catastrophe events on the branches above Ugaritic, with almost no rate heterogeneity elsewhere in the tree; there is a 28\% posterior probability for no catastrophes, compared with a 2\% prior probability.\footnote{To run a Markov chain targeting the prior, we set up a run as usual but tick the box \texttt{Omit traits listed below} and specify \texttt{1:674} to ignore all the traits in the data.}
This indicates that there may be model misfit for Ugaritic.

\begin{figure}
    \centering
    \includegraphics[width=0.7\textwidth, trim=3.15cm 1.5cm 2.85cm 1.5cm, clip=true]{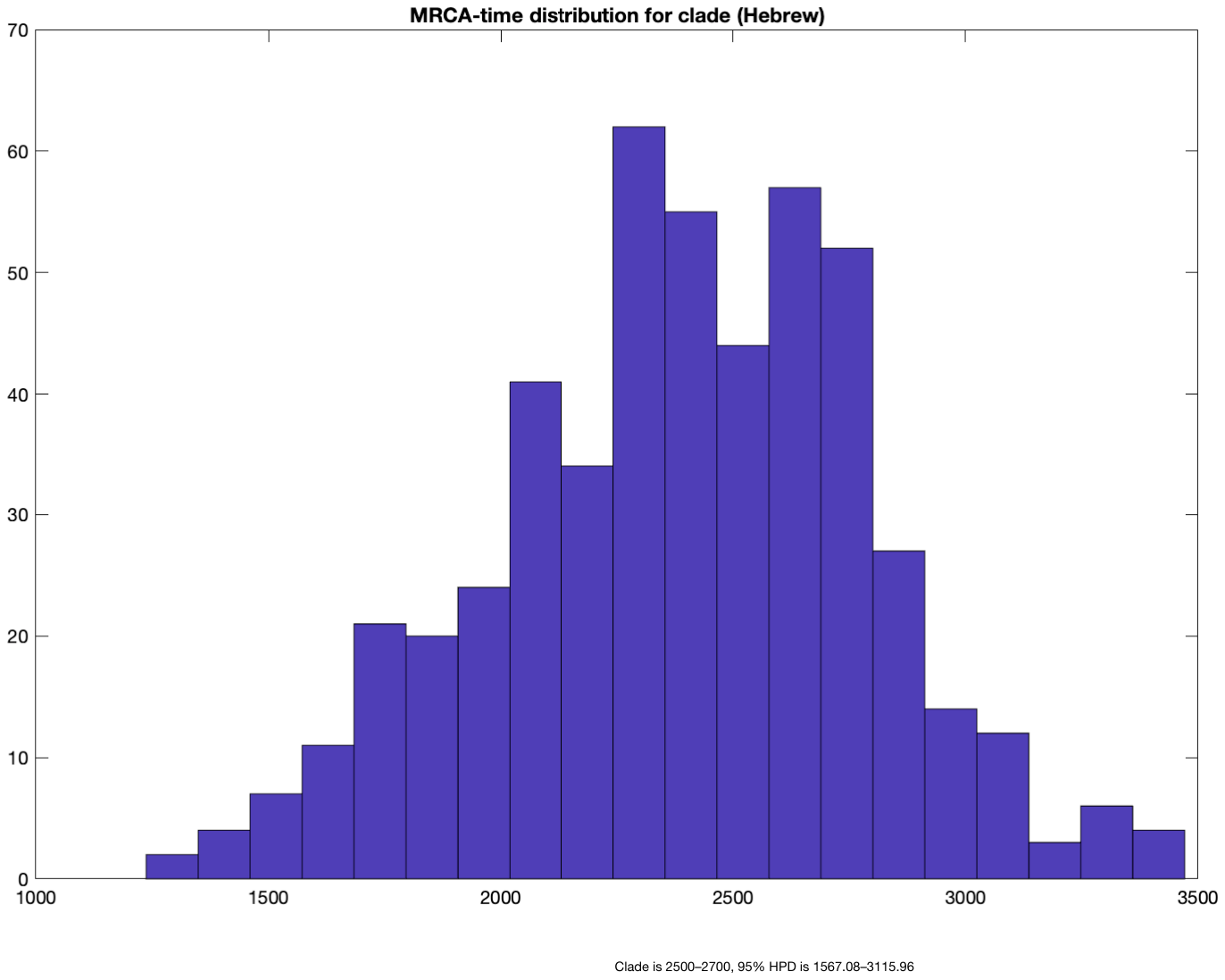}
    \caption{Histogram of samples targeting the posterior distribution on the age of Hebrew after relaxing the clade constraint which restricted it to [2500, 2700].}
    \label{fig:semitic-hebrew}
\end{figure}

We now perform a Bayesian cross-validation analysis of the six leaf age calibration constraints, as discussed by \citet{ryder2011missing}.
To do this, we repeat the MCMC run, with one modification in the settings: in the \texttt{Ignore ages for clades} box, we list clade number 1, for example, then let the MCMC run for a sufficient number of iterations.
Clade number 1 gives constraints on the age and branching time of Hebrew.
With this setting on, these constraints are no longer imposed, letting the Hebrew leaf age and branching time vary freely.
At the end of the run, we use the \texttt{Show age histogram} feature of the Analysis mode to plot the resulting posterior ages, \figref{fig:semitic-hebrew} displays the resulting histogram.
We see that the posterior distribution on ages overlaps with the clade age constraint of [2500, 2700], indicating support for the historically attested constraint.
We repeat this for all time constraints, and find problems with the branching of Biblical Aramaic and the leaf ages of Ugaritic and Gehez.
Given the evidence suggesting that they are outliers, we therefore decide to remove Ugaritic and Gehez from our final analysis.
This improves the model fit, and all remaining constraints are then supported (including Aramaic branching).

\paragraph{Final run}

\begin{figure}
    \includegraphics[width=\textwidth, trim=2cm 0.5cm 3cm 0.5cm, clip=true]{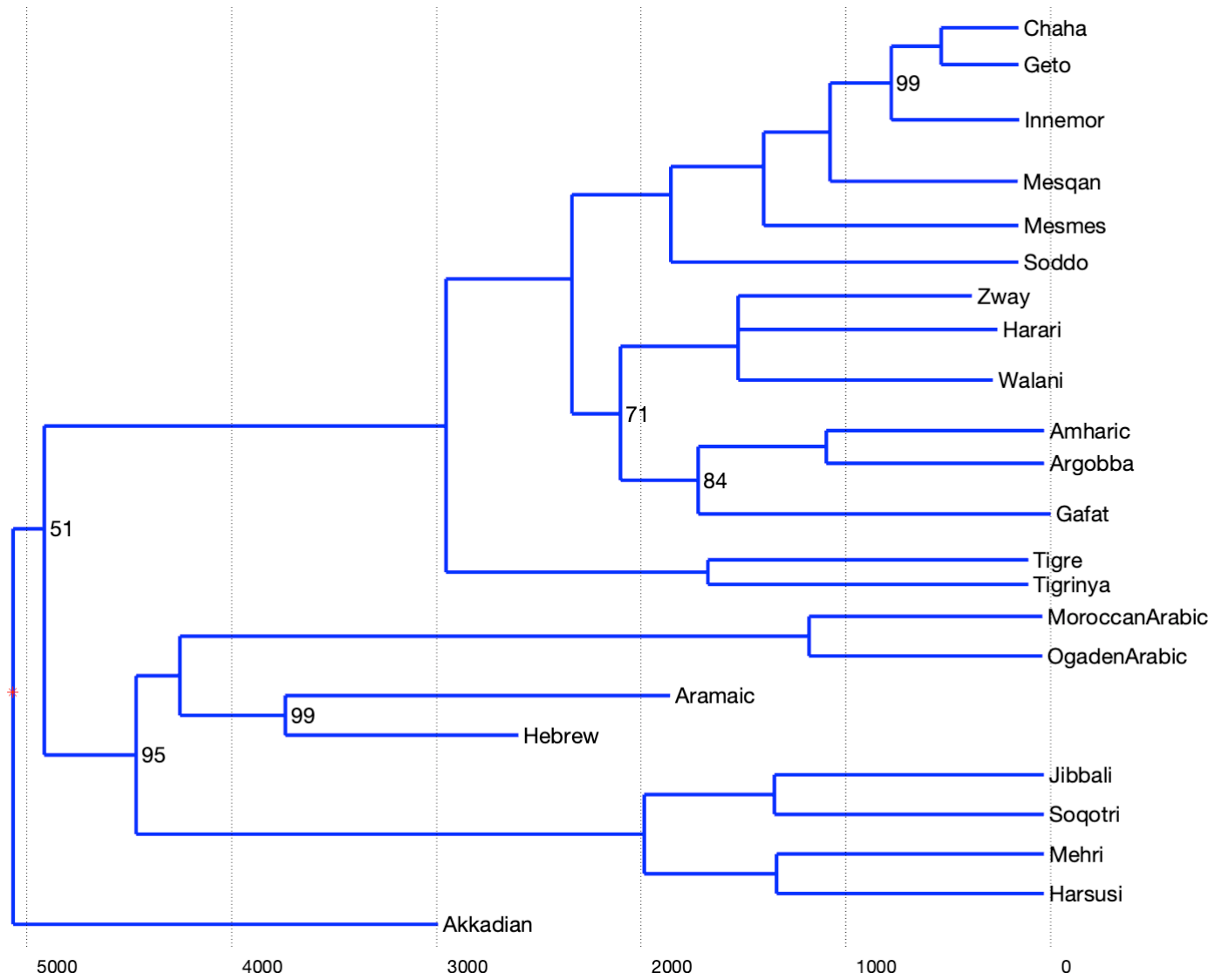}
    \caption{
        Consensus tree with a single catastrophe in the final MCMC run for the Semitic data with Gehez and Ugaritic removed.
    }
    \label{fig:semitic-final-consensus}
\end{figure}

We now perform a final run, from which we will draw our conclusions.
We do not include catastrophes since they seem unnecessary, we also omit taxa 1 and 17 (Gehez and Ugaritic) and ignore clades 2 and 5.
After starting the run, a message appears in the Matlab command window asking whether to delete or ignore Ugaritic from the central clade: we choose to delete it in this case.
With this run, we build the consensus tree displayed in \figref{fig:semitic-final-consensus} and estimate a 95\% highest posterior density interval of [4354, 5624] years Before Present for the root age of the Semitic family.

\end{appendix}

\nocite{gould1970dds}
\bibliographystyle{imsart-nameyear}
\bibliography{references}

\begin{thebibliography}{43}

\bibitem[\protect\citeauthoryear{Alekseyenko, Lee and
  Suchard}{2008}]{alekseyenko2008wad}
\begin{barticle}[author]
\bauthor{\bsnm{Alekseyenko},~\bfnm{A.~V.}\binits{A.~V.}},
  \bauthor{\bsnm{Lee},~\bfnm{C.}\binits{C.}} \AND
  \bauthor{\bsnm{Suchard},~\bfnm{M.}\binits{M.}}
(\byear{2008}).
\btitle{{Wagner and Dollo: A Stochastic Duet by Composing Two Parsimonious
  Solos}}.
\bjournal{Syst. Biol.}
\bvolume{57}
\bpages{772--784}.
\end{barticle}
\endbibitem

\bibitem[\protect\citeauthoryear{Atkinson}{2006}]{atkinson2006species}
\begin{bphdthesis}[author]
\bauthor{\bsnm{Atkinson},~\bfnm{Quentin~Douglas}\binits{Q.~D.}}
(\byear{2006}).
\btitle{From species to languages: a phylogenetic approach to human
  prehistory},
\btype{PhD thesis},
\bpublisher{The University of Auckland}.
\end{bphdthesis}
\endbibitem

\bibitem[\protect\citeauthoryear{Atkinson et~al.}{2005}]{atkinson2005wdw}
\begin{barticle}[author]
\bauthor{\bsnm{Atkinson},~\bfnm{Q.}\binits{Q.}},
  \bauthor{\bsnm{Nicholls},~\bfnm{G.}\binits{G.}},
  \bauthor{\bsnm{Welch},~\bfnm{D.}\binits{D.}} \AND
  \bauthor{\bsnm{Gray},~\bfnm{R.}\binits{R.}}
(\byear{2005}).
\btitle{{From words to dates: water into wine, mathemagic or phylogenetic
  inference?}}
\bjournal{Trans. Philol. Soc.}
\bvolume{103}
\bpages{193--219}.
\end{barticle}
\endbibitem

\bibitem[\protect\citeauthoryear{Barban\c{c}on et~al.}{2013}]{barbancon2013}
\begin{barticle}[author]
\bauthor{\bsnm{Barban\c{c}on},~\bfnm{Fran\c{c}ois}\binits{F.}},
  \bauthor{\bsnm{Evans},~\bfnm{Steven~N.}\binits{S.~N.}},
  \bauthor{\bsnm{Nakhleh},~\bfnm{Luay}\binits{L.}},
  \bauthor{\bsnm{Ringe},~\bfnm{Don}\binits{D.}} \AND
  \bauthor{\bsnm{Warnow},~\bfnm{Tandy}\binits{T.}}
(\byear{2013}).
\btitle{An experimental study comparing linguistic phylogenetic reconstruction
  methods}.
\bjournal{Diachronica}
\bvolume{30}
\bpages{143--170}.
\end{barticle}
\endbibitem

\bibitem[\protect\citeauthoryear{Biswas, Jacob and Vanetti}{2019}]{biswas19}
\begin{binproceedings}[author]
\bauthor{\bsnm{Biswas},~\bfnm{N.}\binits{N.}},
  \bauthor{\bsnm{Jacob},~\bfnm{P.~E.}\binits{P.~E.}} \AND
  \bauthor{\bsnm{Vanetti},~\bfnm{P.}\binits{P.}}
(\byear{2019}).
\btitle{{Estimating convergence of Markov chains with $ L $-lag couplings}}.
In \bbooktitle{NeurIPS}
\bpages{7389--7399}.
\end{binproceedings}
\endbibitem

\bibitem[\protect\citeauthoryear{Bouckaert, Bowern and
  Atkinson}{2018}]{bouckaert2018origin}
\begin{barticle}[author]
\bauthor{\bsnm{Bouckaert},~\bfnm{Remco~R}\binits{R.~R.}},
  \bauthor{\bsnm{Bowern},~\bfnm{Claire}\binits{C.}} \AND
  \bauthor{\bsnm{Atkinson},~\bfnm{Quentin~D}\binits{Q.~D.}}
(\byear{2018}).
\btitle{The origin and expansion of Pama--Nyungan languages across Australia}.
\bjournal{Nature ecology \& evolution}
\bvolume{2}
\bpages{741--749}.
\end{barticle}
\endbibitem

\bibitem[\protect\citeauthoryear{Bouckaert and
  Robbeets}{2017}]{bouckaert2017pseudo}
\begin{barticle}[author]
\bauthor{\bsnm{Bouckaert},~\bfnm{Remco~R}\binits{R.~R.}} \AND
  \bauthor{\bsnm{Robbeets},~\bfnm{Martine}\binits{M.}}
(\byear{2017}).
\btitle{Pseudo Dollo models for the evolution of binary characters along a
  tree}.
\bjournal{BioRxiv}
\bpages{207571}.
\end{barticle}
\endbibitem

\bibitem[\protect\citeauthoryear{Bouckaert et~al.}{2019}]{bouckaert2019beast2}
\begin{barticle}[author]
\bauthor{\bsnm{Bouckaert},~\bfnm{R.}\binits{R.}},
  \bauthor{\bsnm{Vaughan},~\bfnm{T.~G.}\binits{T.~G.}},
  \bauthor{\bsnm{Barido-Sottani},~\bfnm{J.}\binits{J.}},
  \bauthor{\bsnm{Duchêne},~\bfnm{S.}\binits{S.}},
  \bauthor{\bsnm{Fourment},~\bfnm{M.}\binits{M.}},
  \bauthor{\bsnm{Gavryushkina},~\bfnm{A.}\binits{A.}},
  \bauthor{\bsnm{Heled},~\bfnm{J.}\binits{J.}},
  \bauthor{\bsnm{Jones},~\bfnm{G.}\binits{G.}},
  \bauthor{\bsnm{Kühnert},~\bfnm{D.}\binits{D.}},
  \bauthor{\bsnm{De~Maio},~\bfnm{N.}\binits{N.}},
  \bauthor{\bsnm{Matschiner},~\bfnm{M.}\binits{M.}},
  \bauthor{\bsnm{Mendes},~\bfnm{F.~K.}\binits{F.~K.}},
  \bauthor{\bsnm{Müller},~\bfnm{N.~F.}\binits{N.~F.}},
  \bauthor{\bsnm{Ogilvie},~\bfnm{H.~A.}\binits{H.~A.}}, \bauthor{\bparticle{du}
  \bsnm{Plessis},~\bfnm{L.}\binits{L.}},
  \bauthor{\bsnm{Popinga},~\bfnm{A.}\binits{A.}},
  \bauthor{\bsnm{Rambaut},~\bfnm{A.}\binits{A.}},
  \bauthor{\bsnm{Rasmussen},~\bfnm{D.}\binits{D.}},
  \bauthor{\bsnm{Siveroni},~\bfnm{I.}\binits{I.}},
  \bauthor{\bsnm{Suchard},~\bfnm{M.~A.}\binits{M.~A.}},
  \bauthor{\bsnm{Wu},~\bfnm{C.~H.}\binits{C.~H.}},
  \bauthor{\bsnm{Xie},~\bfnm{D.}\binits{D.}},
  \bauthor{\bsnm{Zhang},~\bfnm{C.}\binits{C.}},
  \bauthor{\bsnm{Stadler},~\bfnm{T.}\binits{T.}} \AND
  \bauthor{\bsnm{Drummond},~\bfnm{A.~J.}\binits{A.~J.}}
(\byear{2019}).
\btitle{{BEAST 2.5: An advanced software platform for Bayesian evolutionary
  analysis}}.
\bjournal{PLOS Comput. Biol.}
\bvolume{15}.
\end{barticle}
\endbibitem

\bibitem[\protect\citeauthoryear{Chang et~al.}{2015}]{chang2015ancestry}
\begin{barticle}[author]
\bauthor{\bsnm{Chang},~\bfnm{Will}\binits{W.}},
  \bauthor{\bsnm{Hall},~\bfnm{David}\binits{D.}},
  \bauthor{\bsnm{Cathcart},~\bfnm{Chundra}\binits{C.}} \AND
  \bauthor{\bsnm{Garrett},~\bfnm{Andrew}\binits{A.}}
(\byear{2015}).
\btitle{Ancestry-constrained phylogenetic analysis supports the Indo-European
  steppe hypothesis}.
\bjournal{Language}
\bpages{194--244}.
\end{barticle}
\endbibitem

\bibitem[\protect\citeauthoryear{Currie et~al.}{2013}]{currie2013cultural}
\begin{barticle}[author]
\bauthor{\bsnm{Currie},~\bfnm{Thomas~E}\binits{T.~E.}},
  \bauthor{\bsnm{Meade},~\bfnm{Andrew}\binits{A.}},
  \bauthor{\bsnm{Guillon},~\bfnm{Myrtille}\binits{M.}} \AND
  \bauthor{\bsnm{Mace},~\bfnm{Ruth}\binits{R.}}
(\byear{2013}).
\btitle{Cultural phylogeography of the Bantu Languages of sub-Saharan Africa}.
\bjournal{Proceedings of the Royal Society B: Biological Sciences}
\bvolume{280}
\bpages{20130695}.
\end{barticle}
\endbibitem

\bibitem[\protect\citeauthoryear{Dollo}{1893}]{dollo1893}
\begin{barticle}[author]
\bauthor{\bsnm{Dollo},~\bfnm{L.}\binits{L.}}
(\byear{1893}).
\btitle{{Les Lois de l'\'Evolution}}.
\bjournal{Bulletin de la Soci\'et\'e belge de G\'eologie, de Pal\'eontologie et
  d'Hydrologie}
\bvolume{7}
\bpages{164--166}.
\bnote{Translated in Gould (1970)}.
\end{barticle}
\endbibitem

\bibitem[\protect\citeauthoryear{Drummond et~al.}{2002}]{drummond2002}
\begin{barticle}[author]
\bauthor{\bsnm{Drummond},~\bfnm{A.~J.}\binits{A.~J.}},
  \bauthor{\bsnm{Nicholls},~\bfnm{G.~K.}\binits{G.~K.}},
  \bauthor{\bsnm{Rodrigo},~\bfnm{A.~G.}\binits{A.~G.}} \AND
  \bauthor{\bsnm{Solomon},~\bfnm{W.}\binits{W.}}
(\byear{2002}).
\btitle{{Estimating Mutation Parameters, Population History and Genealogy
  Simultaneously From Temporally Spaced Sequence Data}}.
\bjournal{Genetics}
\bvolume{161}
\bpages{1307--1320}.
\end{barticle}
\endbibitem

\bibitem[\protect\citeauthoryear{Felsenstein}{1981}]{felsenstein1981}
\begin{bbook}[author]
\bauthor{\bsnm{Felsenstein},~\bfnm{J.}\binits{J.}}
(\byear{1981}).
\btitle{{Inferring Phylogenies}}.
\bpublisher{Sinauer Associates Sunderland, Mass., USA}.
\end{bbook}
\endbibitem

\bibitem[\protect\citeauthoryear{Geyer}{1992}]{geyer1992}
\begin{barticle}[author]
\bauthor{\bsnm{Geyer},~\bfnm{Charles~J.}\binits{C.~J.}}
(\byear{1992}).
\btitle{{Practical Markov Chain Monte Carlo}}.
\bjournal{Stat. Sci.}
\bvolume{7}
\bpages{473--483}.
\end{barticle}
\endbibitem

\bibitem[\protect\citeauthoryear{Gould}{1970}]{gould1970dds}
\begin{barticle}[author]
\bauthor{\bsnm{Gould},~\bfnm{S.~J.}\binits{S.~J.}}
(\byear{1970}).
\btitle{{Dollo on Dollo's Law: Irreversibility and the Status of Evolutionary
  Laws}}.
\bjournal{J. Hist. Biol.}
\bvolume{3}
\bpages{189--212}.
\end{barticle}
\endbibitem

\bibitem[\protect\citeauthoryear{Gray and Atkinson}{2003}]{gray2003ltd}
\begin{barticle}[author]
\bauthor{\bsnm{Gray},~\bfnm{R.~D.}\binits{R.~D.}} \AND
  \bauthor{\bsnm{Atkinson},~\bfnm{Q.~D.}\binits{Q.~D.}}
(\byear{2003}).
\btitle{{Language-tree divergence times support the Anatolian theory of
  Indo-European origin}}.
\bjournal{Nature}
\bvolume{426}
\bpages{435--439}.
\end{barticle}
\endbibitem

\bibitem[\protect\citeauthoryear{Gray, Drummond and
  Greenhill}{2009}]{gray2009language}
\begin{barticle}[author]
\bauthor{\bsnm{Gray},~\bfnm{Russell~D}\binits{R.~D.}},
  \bauthor{\bsnm{Drummond},~\bfnm{Alexei~J}\binits{A.~J.}} \AND
  \bauthor{\bsnm{Greenhill},~\bfnm{Simon~J}\binits{S.~J.}}
(\byear{2009}).
\btitle{Language phylogenies reveal expansion pulses and pauses in Pacific
  settlement}.
\bjournal{Science}
\bvolume{323}
\bpages{479--483}.
\end{barticle}
\endbibitem

\bibitem[\protect\citeauthoryear{Greenhill, Currie and
  Gray}{2009}]{greenhill2009does}
\begin{barticle}[author]
\bauthor{\bsnm{Greenhill},~\bfnm{Simon~J}\binits{S.~J.}},
  \bauthor{\bsnm{Currie},~\bfnm{Thomas~E}\binits{T.~E.}} \AND
  \bauthor{\bsnm{Gray},~\bfnm{Russell~D}\binits{R.~D.}}
(\byear{2009}).
\btitle{Does horizontal transmission invalidate cultural phylogenies?}
\bjournal{Proc. R. Soc. B}
\bvolume{276}
\bpages{2299--2306}.
\end{barticle}
\endbibitem

\bibitem[\protect\citeauthoryear{Greenhill, Heggarty and
  Gray}{2020}]{greenhill2020bayesian}
\begin{barticle}[author]
\bauthor{\bsnm{Greenhill},~\bfnm{Simon~J}\binits{S.~J.}},
  \bauthor{\bsnm{Heggarty},~\bfnm{Paul}\binits{P.}} \AND
  \bauthor{\bsnm{Gray},~\bfnm{Russell~D}\binits{R.~D.}}
(\byear{2020}).
\btitle{Bayesian phylolinguistics}.
\bjournal{The handbook of historical linguistics}
\bvolume{2}
\bpages{226--253}.
\end{barticle}
\endbibitem

\bibitem[\protect\citeauthoryear{Hastings}{1970}]{hastings70monte}
\begin{barticle}[author]
\bauthor{\bsnm{Hastings},~\bfnm{W.~K.}\binits{W.~K.}}
(\byear{1970}).
\btitle{Monte {Carlo} sampling methods using {Markov} chains and their
  applications}.
\bjournal{Biometrika}
\bvolume{57}
\bpages{97--109}.
\end{barticle}
\endbibitem

\bibitem[\protect\citeauthoryear{Heggarty et~al.}{2023}]{heggarty2023language}
\begin{barticle}[author]
\bauthor{\bsnm{Heggarty},~\bfnm{Paul}\binits{P.}},
  \bauthor{\bsnm{Anderson},~\bfnm{Cormac}\binits{C.}},
  \bauthor{\bsnm{Scarborough},~\bfnm{Matthew}\binits{M.}},
  \bauthor{\bsnm{King},~\bfnm{Benedict}\binits{B.}},
  \bauthor{\bsnm{Bouckaert},~\bfnm{Remco}\binits{R.}},
  \bauthor{\bsnm{Jocz},~\bfnm{Lechosław}\binits{L.}},
  \bauthor{\bsnm{Kümmel},~\bfnm{Martin~Joachim}\binits{M.~J.}},
  \bauthor{\bsnm{Jügel},~\bfnm{Thomas}\binits{T.}},
  \bauthor{\bsnm{Irslinger},~\bfnm{Britta}\binits{B.}},
  \bauthor{\bsnm{Pooth},~\bfnm{Roland}\binits{R.}},
  \bauthor{\bsnm{Liljegren},~\bfnm{Henrik}\binits{H.}},
  \bauthor{\bsnm{Strand},~\bfnm{Richard~F.}\binits{R.~F.}},
  \bauthor{\bsnm{Haig},~\bfnm{Geoffrey}\binits{G.}},
  \bauthor{\bsnm{Macák},~\bfnm{Martin}\binits{M.}},
  \bauthor{\bsnm{Kim},~\bfnm{Ronald~I.}\binits{R.~I.}},
  \bauthor{\bsnm{Anonby},~\bfnm{Erik}\binits{E.}},
  \bauthor{\bsnm{Pronk},~\bfnm{Tijmen}\binits{T.}},
  \bauthor{\bsnm{Belyaev},~\bfnm{Oleg}\binits{O.}},
  \bauthor{\bsnm{Dewey-Findell},~\bfnm{Tonya~Kim}\binits{T.~K.}},
  \bauthor{\bsnm{Boutilier},~\bfnm{Matthew}\binits{M.}},
  \bauthor{\bsnm{Freiberg},~\bfnm{Cassandra}\binits{C.}},
  \bauthor{\bsnm{Tegethoff},~\bfnm{Robert}\binits{R.}},
  \bauthor{\bsnm{Serangeli},~\bfnm{Matilde}\binits{M.}},
  \bauthor{\bsnm{Liosis},~\bfnm{Nikos}\binits{N.}},
  \bauthor{\bsnm{Stroński},~\bfnm{Krzysztof}\binits{K.}},
  \bauthor{\bsnm{Schulte},~\bfnm{Kim}\binits{K.}},
  \bauthor{\bsnm{Gupta},~\bfnm{Ganesh~Kumar}\binits{G.~K.}},
  \bauthor{\bsnm{Haak},~\bfnm{Wolfgang}\binits{W.}},
  \bauthor{\bsnm{Krause},~\bfnm{Johannes}\binits{J.}},
  \bauthor{\bsnm{Atkinson},~\bfnm{Quentin~D.}\binits{Q.~D.}},
  \bauthor{\bsnm{Greenhill},~\bfnm{Simon~J.}\binits{S.~J.}},
  \bauthor{\bsnm{Kühnert},~\bfnm{Denise}\binits{D.}} \AND
  \bauthor{\bsnm{Gray},~\bfnm{Russell~D.}\binits{R.~D.}}
(\byear{2023}).
\btitle{{Language trees with sampled ancestors support a hybrid model for the
  origin of Indo-European languages}}.
\bjournal{Science}
\bvolume{381}
\bpages{eabg0818}.
\end{barticle}
\endbibitem

\bibitem[\protect\citeauthoryear{Jacob, O’Leary and Atchadé}{2020}]{jacob20}
\begin{barticle}[author]
\bauthor{\bsnm{Jacob},~\bfnm{P.~E.}\binits{P.~E.}},
  \bauthor{\bsnm{O’Leary},~\bfnm{J.}\binits{J.}} \AND
  \bauthor{\bsnm{Atchadé},~\bfnm{Y.~F.}\binits{Y.~F.}}
(\byear{2020}).
\btitle{{Unbiased Markov chain Monte Carlo methods with couplings}}.
\bjournal{J. Roy. Statist. Soc. B}
\bvolume{82}
\bpages{543--600}.
\end{barticle}
\endbibitem

\bibitem[\protect\citeauthoryear{Kelly}{2016}]{kelly2016dphil}
\begin{bphdthesis}[author]
\bauthor{\bsnm{Kelly},~\bfnm{L.~J.}\binits{L.~J.}}
(\byear{2016}).
\btitle{{A Stochastic Dollo model for lateral transfer}},
\btype{PhD thesis},
\bpublisher{University of Oxford}.
\end{bphdthesis}
\endbibitem

\bibitem[\protect\citeauthoryear{{Kelly} and
  {Nicholls}}{2017}]{kelly2017lateral}
\begin{barticle}[author]
\bauthor{\bsnm{{Kelly}},~\bfnm{L.~J.}\binits{L.~J.}} \AND
  \bauthor{\bsnm{{Nicholls}},~\bfnm{G.~K.}\binits{G.~K.}}
(\byear{2017}).
\btitle{{Lateral transfer in Stochastic Dollo models}}.
\bjournal{Ann. Appl. Stat.}
\bvolume{11}
\bpages{1146--1168}.
\end{barticle}
\endbibitem

\bibitem[\protect\citeauthoryear{Kelly, Ryder and
  Clart{\'e}}{2023}]{kelly2023lagged}
\begin{barticle}[author]
\bauthor{\bsnm{Kelly},~\bfnm{L.~J.}\binits{L.~J.}},
  \bauthor{\bsnm{Ryder},~\bfnm{R.~J.}\binits{R.~J.}} \AND
  \bauthor{\bsnm{Clart{\'e}},~\bfnm{G.}\binits{G.}}
(\byear{2023}).
\btitle{{Lagged couplings diagnose Markov chain Monte Carlo phylogenetic
  inference}}.
\bjournal{Ann. Appl. Stat.}
\bvolume{17}
\bpages{1419--1443}.
\end{barticle}
\endbibitem

\bibitem[\protect\citeauthoryear{Kitchen et~al.}{2009}]{kitchen2009bayesian}
\begin{barticle}[author]
\bauthor{\bsnm{Kitchen},~\bfnm{Andrew}\binits{A.}},
  \bauthor{\bsnm{Ehret},~\bfnm{Christopher}\binits{C.}},
  \bauthor{\bsnm{Assefa},~\bfnm{Shiferaw}\binits{S.}} \AND
  \bauthor{\bsnm{Mulligan},~\bfnm{Connie~J}\binits{C.~J.}}
(\byear{2009}).
\btitle{Bayesian Phylogenetic Analysis of Semitic Languages Identifies an Early
  Bronze Age Origin of Semitic in the Near East}.
\bjournal{Proc. R. Soc. B}
\bvolume{276}
\bpages{2703--2710}.
\end{barticle}
\endbibitem

\bibitem[\protect\citeauthoryear{Koch}{2016}]{Koch2016}
\begin{bbook}[author]
\bauthor{\bsnm{Koch},~\bfnm{Marlies}\binits{M.}}
(\byear{2016}).
\btitle{{Geschichte der gesprochenen Sprache von Bayerisch-Schwaben
  Phonologische Untersuchungen mittels diatopisch orientierter
  Rekonstruktion}}.
\bpublisher{BiblioScout}.
\bdoi{10.25162/9783515114035}
\end{bbook}
\endbibitem

\bibitem[\protect\citeauthoryear{Koile et~al.}{2022}]{koile2022phylogeographic}
\begin{barticle}[author]
\bauthor{\bsnm{Koile},~\bfnm{Ezequiel}\binits{E.}},
  \bauthor{\bsnm{Greenhill},~\bfnm{Simon~J}\binits{S.~J.}},
  \bauthor{\bsnm{Blasi},~\bfnm{Dami{\'a}n~E}\binits{D.~E.}},
  \bauthor{\bsnm{Bouckaert},~\bfnm{Remco}\binits{R.}} \AND
  \bauthor{\bsnm{Gray},~\bfnm{Russell~D}\binits{R.~D.}}
(\byear{2022}).
\btitle{Phylogeographic analysis of the Bantu language expansion supports a
  rainforest route}.
\bjournal{Proc. Natl. Acad. Sci. U.S.A.}
\bvolume{119}
\bpages{e2112853119}.
\end{barticle}
\endbibitem

\bibitem[\protect\citeauthoryear{Lewis}{2001}]{lewis2001lae}
\begin{barticle}[author]
\bauthor{\bsnm{Lewis},~\bfnm{P.~O.}\binits{P.~O.}}
(\byear{2001}).
\btitle{{A Likelihood Approach to Estimating Phylogeny from Discrete
  Morphological Character Data}}.
\bjournal{Syst. Biol.}
\bvolume{50}
\bpages{913--925}.
\end{barticle}
\endbibitem

\bibitem[\protect\citeauthoryear{Maddison, Swofford and
  Maddison}{1997}]{maddison1997}
\begin{barticle}[author]
\bauthor{\bsnm{Maddison},~\bfnm{D.~R.}\binits{D.~R.}},
  \bauthor{\bsnm{Swofford},~\bfnm{D.~L.}\binits{D.~L.}} \AND
  \bauthor{\bsnm{Maddison},~\bfnm{W.~P.}\binits{W.~P.}}
(\byear{1997}).
\btitle{Nexus: An Extensible File Format for Systematic Information}.
\bjournal{Syst. Biol.}
\bvolume{46}
\bpages{590--621}.
\end{barticle}
\endbibitem

\bibitem[\protect\citeauthoryear{McPherson
  et~al.}{2016}]{mcpherson2016divergent}
\begin{barticle}[author]
\bauthor{\bsnm{McPherson},~\bfnm{A.}\binits{A.}},
  \bauthor{\bsnm{Roth},~\bfnm{A.}\binits{A.}},
  \bauthor{\bsnm{Laks},~\bfnm{E.}\binits{E.}},
  \bauthor{\bsnm{Masud},~\bfnm{T.}\binits{T.}},
  \bauthor{\bsnm{Bashashati},~\bfnm{A.}\binits{A.}},
  \bauthor{\bsnm{Zhang},~\bfnm{A.~W.}\binits{A.~W.}},
  \bauthor{\bsnm{Ha},~\bfnm{G.}\binits{G.}},
  \bauthor{\bsnm{Biele},~\bfnm{J.}\binits{J.}},
  \bauthor{\bsnm{Yap},~\bfnm{D.}\binits{D.}},
  \bauthor{\bsnm{Wan},~\bfnm{A.}\binits{A.}},
  \bauthor{\bsnm{Prentice},~\bfnm{L.~M.}\binits{L.~M.}},
  \bauthor{\bsnm{Khattra},~\bfnm{J.}\binits{J.}},
  \bauthor{\bsnm{Smith},~\bfnm{M.~A.}\binits{M.~A.}},
  \bauthor{\bsnm{Nielsen},~\bfnm{C.~B.}\binits{C.~B.}},
  \bauthor{\bsnm{Mullaly},~\bfnm{S.~C.}\binits{S.~C.}},
  \bauthor{\bsnm{Kalloger},~\bfnm{S.}\binits{S.}},
  \bauthor{\bsnm{Karnezis},~\bfnm{A.}\binits{A.}},
  \bauthor{\bsnm{Shumansky},~\bfnm{K.}\binits{K.}},
  \bauthor{\bsnm{Siu},~\bfnm{C.}\binits{C.}},
  \bauthor{\bsnm{Rosner},~\bfnm{J.}\binits{J.}},
  \bauthor{\bsnm{Chan},~\bfnm{H.~L.}\binits{H.~L.}},
  \bauthor{\bsnm{Ho},~\bfnm{J.}\binits{J.}},
  \bauthor{\bsnm{Melnyk},~\bfnm{N.}\binits{N.}},
  \bauthor{\bsnm{Senz},~\bfnm{J.}\binits{J.}},
  \bauthor{\bsnm{Yang},~\bfnm{W.}\binits{W.}},
  \bauthor{\bsnm{Moore},~\bfnm{R.}\binits{R.}},
  \bauthor{\bsnm{Mungall},~\bfnm{A.~J.}\binits{A.~J.}},
  \bauthor{\bsnm{Marra},~\bfnm{M.~A.}\binits{M.~A.}},
  \bauthor{\bsnm{Bouchard-C\^{o}t\'{e}},~\bfnm{A.}\binits{A.}},
  \bauthor{\bsnm{Gilks},~\bfnm{C.~B.}\binits{C.~B.}},
  \bauthor{\bsnm{Huntsman},~\bfnm{D.~G}\binits{D.~G.}},
  \bauthor{\bsnm{McAlpine},~\bfnm{J.~N}\binits{J.~N.}},
  \bauthor{\bsnm{Aparicio},~\bfnm{S.}\binits{S.}} \AND
  \bauthor{\bsnm{Shah},~\bfnm{S.~P.}\binits{S.~P.}}
(\byear{2016}).
\btitle{Divergent modes of clonal spread and intraperitoneal mixing in
  high-grade serous ovarian cancer}.
\bjournal{Nature Genet.}
\bvolume{48}
\bpages{758--767}.
\end{barticle}
\endbibitem

\bibitem[\protect\citeauthoryear{Metropolis
  et~al.}{1953}]{metropolis1953equation}
\begin{barticle}[author]
\bauthor{\bsnm{Metropolis},~\bfnm{N.}\binits{N.}},
  \bauthor{\bsnm{Rosenbluth},~\bfnm{A.~W.}\binits{A.~W.}},
  \bauthor{\bsnm{Rosenbluth},~\bfnm{M.~N.}\binits{M.~N.}},
  \bauthor{\bsnm{Teller},~\bfnm{A.~H.}\binits{A.~H.}} \AND
  \bauthor{\bsnm{Teller},~\bfnm{E.}\binits{E.}}
(\byear{1953}).
\btitle{Equation of State Calculations by Fast Computing Machines}.
\bjournal{J. Chem. Phys.}
\bvolume{21}
\bpages{1087--1092}.
\end{barticle}
\endbibitem

\bibitem[\protect\citeauthoryear{Neureiter
  et~al.}{2022}]{neureiter2022detecting}
\begin{barticle}[author]
\bauthor{\bsnm{Neureiter},~\bfnm{Nico}\binits{N.}},
  \bauthor{\bsnm{Ranacher},~\bfnm{Peter}\binits{P.}},
  \bauthor{\bsnm{Efrat-Kowalsky},~\bfnm{Nour}\binits{N.}},
  \bauthor{\bsnm{Kaiping},~\bfnm{Gereon~A}\binits{G.~A.}},
  \bauthor{\bsnm{Weibel},~\bfnm{Robert}\binits{R.}},
  \bauthor{\bsnm{Widmer},~\bfnm{Paul}\binits{P.}} \AND
  \bauthor{\bsnm{Bouckaert},~\bfnm{Remco~R}\binits{R.~R.}}
(\byear{2022}).
\btitle{Detecting contact in language trees: a Bayesian phylogenetic model with
  horizontal transfer}.
\bjournal{Humanit. Soc. Sci. Commun.}
\bvolume{9}
\bpages{1--14}.
\end{barticle}
\endbibitem

\bibitem[\protect\citeauthoryear{Nicholls and Gray}{{2008}}]{nicholls2008}
\begin{barticle}[author]
\bauthor{\bsnm{Nicholls},~\bfnm{Geoff~K.}\binits{G.~K.}} \AND
  \bauthor{\bsnm{Gray},~\bfnm{Russell~D.}\binits{R.~D.}}
(\byear{{2008}}).
\btitle{{Dated Ancestral Trees from Binary Trait Data and its Application to
  the Diversification of Languages}}.
\bjournal{{J. Roy. Statist. Soc. B}}
\bvolume{70}
\bpages{545--566}.
\end{barticle}
\endbibitem

\bibitem[\protect\citeauthoryear{Nicholls and
  Ryder}{2011}]{nicholls2011phylogenetic}
\begin{binproceedings}[author]
\bauthor{\bsnm{Nicholls},~\bfnm{Geoff~K}\binits{G.~K.}} \AND
  \bauthor{\bsnm{Ryder},~\bfnm{Robin~J}\binits{R.~J.}}
(\byear{2011}).
\btitle{Phylogenetic Models for Semitic Vocabulary}.
In \bbooktitle{Proceedings of the International Workshop on Statistical
  Modelling}
(\beditor{\bfnm{D.}\binits{D.}~\bsnm{Conesa}},
  \beditor{\bfnm{A.}\binits{A.}~\bsnm{Forte}},
  \beditor{\bfnm{A.}\binits{A.}~\bsnm{L\'{o}pez-Qu\'{\i}lez}} \AND
  \beditor{\bfnm{F.}\binits{F.}~\bsnm{Mu\~{n}oz}}, eds.)
\bpages{431--436}.
\end{binproceedings}
\endbibitem

\bibitem[\protect\citeauthoryear{Pagel, Meade and
  Barker}{2004}]{pagel2004bayesian}
\begin{barticle}[author]
\bauthor{\bsnm{Pagel},~\bfnm{M.}\binits{M.}},
  \bauthor{\bsnm{Meade},~\bfnm{A.}\binits{A.}} \AND
  \bauthor{\bsnm{Barker},~\bfnm{D.}\binits{D.}}
(\byear{2004}).
\btitle{{Bayesian estimation of ancestral character states on phylogenies}}.
\bjournal{Syst. Biol.}
\bvolume{53}
\bpages{673--684}.
\end{barticle}
\endbibitem

\bibitem[\protect\citeauthoryear{Rambaut et~al.}{2018}]{rambaut18tracer}
\begin{barticle}[author]
\bauthor{\bsnm{Rambaut},~\bfnm{A.}\binits{A.}},
  \bauthor{\bsnm{Drummond},~\bfnm{A.~J.}\binits{A.~J.}},
  \bauthor{\bsnm{Xie},~\bfnm{D.}\binits{D.}},
  \bauthor{\bsnm{Baele},~\bfnm{G.}\binits{G.}} \AND
  \bauthor{\bsnm{Suchard},~\bfnm{M.~A.}\binits{M.~A.}}
(\byear{2018}).
\btitle{{Posterior Summarization in Bayesian Phylogenetics Using Tracer 1.7}}.
\bjournal{Syst. Biol.}
\bvolume{67}
\bpages{901--904}.
\end{barticle}
\endbibitem

\bibitem[\protect\citeauthoryear{Ronquist et~al.}{2012}]{ronquist12mrbayes}
\begin{barticle}[author]
\bauthor{\bsnm{Ronquist},~\bfnm{F.}\binits{F.}},
  \bauthor{\bsnm{Teslenko},~\bfnm{M.}\binits{M.}}, \bauthor{\bsnm{Van
  Der~Mark},~\bfnm{P.}\binits{P.}},
  \bauthor{\bsnm{Ayres},~\bfnm{D.~L.}\binits{D.~L.}},
  \bauthor{\bsnm{Darling},~\bfnm{A.}\binits{A.}},
  \bauthor{\bsnm{H{\"o}hna},~\bfnm{S.}\binits{S.}},
  \bauthor{\bsnm{Larget},~\bfnm{B.}\binits{B.}},
  \bauthor{\bsnm{Liu},~\bfnm{L.}\binits{L.}},
  \bauthor{\bsnm{Suchard},~\bfnm{M.~A.}\binits{M.~A.}} \AND
  \bauthor{\bsnm{Huelsenbeck},~\bfnm{J.~P.}\binits{J.~P.}}
(\byear{2012}).
\btitle{{MrBayes 3.2: efficient Bayesian phylogenetic inference and model
  choice across a large model space}}.
\bjournal{Syst. Biol.}
\bvolume{61}
\bpages{539--542}.
\end{barticle}
\endbibitem

\bibitem[\protect\citeauthoryear{Ryder}{2010}]{ryder2010dphil}
\begin{bphdthesis}[author]
\bauthor{\bsnm{Ryder},~\bfnm{Robin~J.}\binits{R.~J.}}
(\byear{2010}).
\btitle{{Phylogenetic Models of Language Diversification}},
\btype{PhD thesis},
\bpublisher{{University of Oxford}}.
\end{bphdthesis}
\endbibitem

\bibitem[\protect\citeauthoryear{Ryder and Nicholls}{2011}]{ryder2011missing}
\begin{barticle}[author]
\bauthor{\bsnm{Ryder},~\bfnm{Robin~J.}\binits{R.~J.}} \AND
  \bauthor{\bsnm{Nicholls},~\bfnm{Geoff~K.}\binits{G.~K.}}
(\byear{2011}).
\btitle{{Missing Data in a Stochastic Dollo Model for Binary Trait Data, and
  its Application to the Dating of Proto-Indo-European}}.
\bjournal{J. Roy. Statist. Soc. C}.
\end{barticle}
\endbibitem

\bibitem[\protect\citeauthoryear{Sagart et~al.}{2019}]{sagart2019dated}
\begin{barticle}[author]
\bauthor{\bsnm{Sagart},~\bfnm{Laurent}\binits{L.}},
  \bauthor{\bsnm{Jacques},~\bfnm{Guillaume}\binits{G.}},
  \bauthor{\bsnm{Lai},~\bfnm{Yunfan}\binits{Y.}},
  \bauthor{\bsnm{Ryder},~\bfnm{Robin~J}\binits{R.~J.}},
  \bauthor{\bsnm{Thouzeau},~\bfnm{Valentin}\binits{V.}},
  \bauthor{\bsnm{Greenhill},~\bfnm{Simon~J}\binits{S.~J.}} \AND
  \bauthor{\bsnm{List},~\bfnm{Johann-Mattis}\binits{J.-M.}}
(\byear{2019}).
\btitle{Dated language phylogenies shed light on the ancestry of Sino-Tibetan}.
\bjournal{Proc. Natl. Acad. Sci. U.S.A.}
\bvolume{116}
\bpages{10317--10322}.
\end{barticle}
\endbibitem

\bibitem[\protect\citeauthoryear{Suchard et~al.}{2018}]{suchard2018beast1}
\begin{barticle}[author]
\bauthor{\bsnm{Suchard},~\bfnm{M.~A.}\binits{M.~A.}},
  \bauthor{\bsnm{Lemey},~\bfnm{P.}\binits{P.}},
  \bauthor{\bsnm{Baele},~\bfnm{G.}\binits{G.}},
  \bauthor{\bsnm{Ayres},~\bfnm{D.~L.}\binits{D.~L.}},
  \bauthor{\bsnm{Drummond},~\bfnm{A.~J.}\binits{A.~J.}} \AND
  \bauthor{\bsnm{Rambaut},~\bfnm{A.}\binits{A.}}
(\byear{2018}).
\btitle{{Bayesian phylogenetic and phylodynamic data integration using BEAST
  1.10}}.
\bjournal{Virus Evol.}
\bvolume{4}.
\end{barticle}
\endbibitem

\bibitem[\protect\citeauthoryear{Thomson et~al.}{2014}]{thomson2014critical}
\begin{barticle}[author]
\bauthor{\bsnm{Thomson},~\bfnm{R.~C.}\binits{R.~C.}},
  \bauthor{\bsnm{Plachetzki},~\bfnm{D.~C.}\binits{D.~C.}},
  \bauthor{\bsnm{Mahler},~\bfnm{D.~L.}\binits{D.~L.}} \AND
  \bauthor{\bsnm{Moore},~\bfnm{B.~R.}\binits{B.~R.}}
(\byear{2014}).
\btitle{{A critical appraisal of the use of microRNA data in phylogenetics}}.
\bjournal{Proc. Natl. Acad. Sci. U.S.A.}
\bvolume{111}
\bpages{E3659-E3668}.
\end{barticle}
\endbibitem

\end{thebibliography}

\end{document}